\documentclass[11pt]{article}

\usepackage[section]{algorithm}
\usepackage{algpseudocode}
\usepackage{amsmath}
\usepackage{amssymb}
\usepackage{amsthm}
\usepackage{authblk}
\usepackage{bm}
\usepackage{booktabs}
\usepackage[strict]{changepage}
\usepackage{enumitem}
\usepackage[a4paper,bottom=3cm]{geometry}
\usepackage[utf8]{inputenc}
\usepackage{listings}
\usepackage{mathtools}
\usepackage{microtype}
\usepackage{siunitx}
\usepackage{stmaryrd}
\usepackage{xifthen}
\usepackage{xspace}
\usepackage[hyperref]{xcolor}

\usepackage{hyperref}

\lstdefinelanguage[Dune]{C++}[]{C++}{
  morekeywords=[1]{
     default_random_engine,
     tie,
     uniform_int_distribution
  }
}

\lstdefinestyle{float}{
  basicstyle=\footnotesize\ttfamily,
  commentstyle=\color{black!75}\ttfamily,
  stringstyle=\ttfamily,
  backgroundcolor=\color{blue!5},
  frame=lines,
  numbers=left,
  numberstyle=\tiny,
  keywordstyle=\color{blue!75!black}
}

\lstdefinestyle{display}{
  basicstyle=\footnotesize\ttfamily,
  numbers=left,
  numberstyle=\tiny,
  keywordstyle=\color{blue!75!black},
  xleftmargin=1em
}

\newcounter{steps}
\newcounter{step}

\newenvironment{steps}{%
  \begin{adjustwidth}{1em}{}
  \begin{description}[style=unboxed, leftmargin=0em,font=\normalfont\emph]%
  \let\olditem\item%
  \renewcommand\item[1][]{%
    \refstepcounter{step}%
    \ifthenelse{\isempty{##1}}{
      \olditem[\textbf{Step \thestep.}]%
    }{
      \olditem[\textbf{Step \thestep~(##1).}]%
    }
  }%
}{
  \let\item\olditem%
  \end{description}
  \end{adjustwidth}
  \setcounter{step}{0}%
  \stepcounter{steps}%
}

\newtheorem{definition}{Definition}[section]
\newtheorem{example}[definition]{Example}
\newtheorem{remark}[definition]{Remark}

\newcommand{\dune}[1][]{%
  \ifthenelse{\isempty{#1}}{%
    \textsc{Dune}\xspace%
  }{%
    \textsc{Dune-{#1}}\xspace%
  }%
}

\newcommand{\entities}[1][]{%
  \ifthenelse{\isempty{#1}}{%
    \mathcal{E}
  }{%
    \mathcal{E}^{#1}
  }%
}

\renewcommand{\phi}{\varphi}

\newcommand{\N}{\mathbb{N}}
\newcommand{\R}{\mathbb{R}}

\newcommand{\abs}[1]{\lvert #1 \rvert}
\newcommand{\jump}[1]{\llbracket #1 \rrbracket}
\newcommand{\mean}[1]{\{#1\}}
\newcommand{\norm}[1]{\lVert #1 \rVert}
\newcommand{\restr}[1]{\mathord{\restriction}_{#1}}

\newcommand{\basis}{\mathcal{B}}
\newcommand{\dofs}{\mathcal{D}}
\newcommand{\grid}{\mathcal{G}}
\newcommand{\intersections}[2][]{
  \ifthenelse{\isempty{#1}}{%
    \mathcal{I}(#2)
  }{%
    \mathcal{I}_{#1}(#2)
  }%
}

\newcommand*\dd{\mathop{}\!\mathrm{d}}
\newcommand{\Cpp}{C++}

\title{Implementation of \textit{hp}-adaptive discontinuous finite element
methods in \textsc{Dune-Fem}}

\author{Christoph Gersbacher}

\affil{\small %
  Department of Applied Mathematics, University of Freiburg, \\ %
  Hermann-Herder-Str. 10, D-79104 Freiburg, Germany. %
}

\date{April, 2016}

\begin{document}

\maketitle


\begin{abstract}
In this paper we describe generic algorithms and data structures for the
implementation of $hp$-adaptive discontinuous finite element methods in the
\dune[Fem] library. Special attention is given to the often tedious and
error-prone task of transferring user data during adaptation.
Simultaneously, we generalize the approach to the restriction and
prolongation of data currently implemented in \dune[Fem] to the case of
$p$- and $hp$-adaptation. The \texttt{dune-fem-hpdg} module described in
this paper provides an extensible reference implementation of $hp$-adaptive
discontinuous discrete function spaces. We give details on its
implementation and the extended adaptive interface. As proof of concept we
present the practical realization of an $hp$-adaptive interior penalty
method for elliptic problems.
\end{abstract}


\section{Introduction}

Adaptive mesh refinement or $h$-refinement is today a standard tool in the
acceleration of finite element methods. The $hp$-version of the finite
element method was made popular by Babu{\v{s}}ka and co-workers in a series
of papers \cite{Babuska1981, Gui1986, Gui1986a, Gui1986b, Guo1986,
Guo1986a}. They showed that in case the solution is sufficiently smooth an
adaptation in $p$, i.e., in the polynomial order of a local approximation
space, may be more advantageous. By a proper combination of $h$- and
$p$-refinement exponential convergence towards the exact solution in terms
of the number of degrees of freedom may be observed.

From a theoretical point of view $hp$-adaptive finite element methods are a
natural generalization of $h$-adaptive methods. However, their practical
implementation still poses considerable difficulties and few open source
software solutions are available to researchers. Of these we want to
mention the \Cpp libraries
  \emph{Concepts}
    \cite{Lage1998, Frauenfelder2002}
and
  \emph{deal.II}
    \cite{Bangerth2007, Bangerth2009},
and the Fortran 90 library
  \emph{PHAML}
    \cite{Mitchell2006}.

In the present paper we want to discuss data structures and algorithms for
implementing $hp$-adaptive discontinuous finite element methods in the
\dune[Fem] finite element library \cite{Dedner2010}. The main difficulty
we want to address is the restriction and prolongation of user data in
adaptive computations. In case of local mesh refinement utility classes in
\dune[Fem] relive users of this often tedious and error-prone task. The
dedicated data structures and algorithms need to be generalized to the case
of $p$- and $hp$-refinement in such a way that legacy code remains valid.

At the time of writing there has been one prior attempt to implementing
$hp$-adaptive finite element methods in \dune[Fem]. In a pioneering work
Andreas Dedner and Robert Kl{\"o}fkorn explored how to implement continuous
and discontinuous $hp$-adaptive finite element spaces in \dune[Fem]. For
the most part, the restriction and prolongation of discrete functions may
be considered as beyond the scope of this implementation and was done in a
preliminary way only. For the local approximation Lagrange ansatz
polynomials are used. The software was used in the preparation of numerical
results for stationary problems in \cite{Dedner2014}.

This paper is accompanied by a new extension module to the \dune[Fem]
library. The \texttt{dune-fem-hpdg} module provides a number of discrete
function spaces for implementing $p$- and $hp$-adaptive discontinuous
finite element methods in \dune[Fem]. Hardly any restrictions to the local
ansatz functions are imposed. The software is extensible and allows users
to quickly setup new finite element spaces using customized ansatz
functions with minimal effort. The restriction and prolongation of discrete
functions and other user data is handled in a fully automated fashion. The
software is compatible with \dune[Fem] 2.4 and in particular with its data
structures and algorithms for local mesh refinement.

The outline of this paper is as follows. In \mbox{Section
\ref{sct:discrete_function_spaces}} we briefly revisit the abstraction
principles behind the \dune[Fem] finite element library. \mbox{Section
\ref{sct:data_transfer}} will be concerned with the restriction and
prolongation of user data in $hp$- adaptive simulations. In \mbox{Section
\ref{sct:dune_fem_hpdg}} we describe the implementation and usage of the
\texttt{dune-fem-hpdg} module. As proof of concept we implemented an
$hp$-adaptive interior penalty method for an elliptic model problem, the
numerical results are shown in \mbox{Section \ref{sct:numerical_results}}.


\section{Discrete functions and discrete function spaces}
\label{sct:discrete_function_spaces}

Let $\Omega \subset \R^d$ be a bounded domain. We assume that the
computational domain is discretized by a grid
\begin{math}
  \grid = \{ E_i \subset \R^d \mid i \in I \},
\end{math}
i.e., by a finite number of closed grid cells with non-overlapping
interiors such that
\begin{equation*}
  \bigcup_{E \in \grid} E = \overline{\Omega}.
\end{equation*}
A \emph{discontinuous finite element space} or \emph{discontinuous Galerkin
space} is a finite-dimensional, piecewise continuous function space
\begin{equation*}
  X(\grid) = \left\{ u \in L^\infty(\Omega)
    \mid u\restr{E} \in X(E) \text{ for all } E \in \grid \right\},
\end{equation*}
where for each $E \in \grid$ we fixed a finite-dimensional subspace, the
so-called local space $X(E) \subset C(E)$. For discretization purposes we
need to fix a basis for each local space.

\begin{definition}[Local basis function set]
Let $X(E)$ be a local space with $\dim(X(E)) = n_E$. A basis
\begin{equation*}
  \basis_E = \left\{ \phi_{E,i} \mid i = 0, \ldots, n_E-1 \right\}
\end{equation*}
of $X(E)$ is called a \emph{local basis function set}.
\end{definition}

A function $u \in X(\grid)$ is called a \emph{discrete function}. Note that
on inter-element intersections a discrete function is double-valued. The
restriction of $u$ to a grid element $E \in \grid$ is called a \emph{local
(discrete) function}. In the fixed local basis $\basis_E$ a local function
reads as follows,
\begin{equation*}
  u\restr{E} = \sum_{i = 0}^{n_E-1} u_{E,i} \phi_{E,i}.
\end{equation*}
The scalars
\begin{math}
  u_{E,i} \in \R,~i = 0, \ldots, n_E-1,
\end{math}
are called \emph{local degrees of freedom} (DOFs); the vector
\begin{math}
  \left(u_{E,i}\right)_{i = 0, \ldots, n_E-1}
\end{math}
is called the \emph{local DOF vector}. For practical purposes we want to
store the vector of all DOFs in a consecutive array of floating point
numbers. This requires a suitable enumeration of the DOFs.

\begin{definition}[Local DOF mapping]
Let $N = \dim(X(\grid))$ denote the dimension of the finite element space
$X(\grid)$. A \emph{local DOF mapping} for $E \in \grid$ is an injective
mapping
\begin{equation*}
  \mu_E: \{0, \ldots, n_E-1\} \rightarrow \{0, \ldots, N-1\}
\end{equation*}
from a local DOF enumeration to global indices.
\end{definition}

Having fixed families of local basis function sets and local DOF mappings
we have  at the same time fixed a basis of $X(\grid)$.

\begin{definition}[Global basis function set]
\label{dfn:global_basis}
Let
\begin{math}
  \left(\basis_E\right)_{E \in \grid}
\end{math}
and
\begin{math}
  \left(\mu_E\right)_{E \in \grid}
\end{math}
be fixed families of local basis function sets and DOF mappings. Define
\begin{math}
  \psi_i \in X(\grid),~i = 1, \ldots, N-1,
\end{math}
by
\begin{equation*}
  \psi_i \restr{E} = \begin{cases}
      \phi_{E,j} & \text{if } \mu_E(j) = i, \\
      0 & \text{otherwise}
    \end{cases}
  \qquad (E \in \grid).
\end{equation*}
Then, the set
\begin{math}
  \basis_\grid = \left\{ \psi_i \mid i = 0, \ldots, N-1 \right\}
\end{math}
forms a basis of $X(\grid)$.
\end{definition}

The following definition is a slightly modified version of
\cite[Definition 5]{Dedner2010} which represents the theoretical
foundation for the implementations of finite element spaces spaces in
\dune[Fem].

\begin{definition}[Discrete function space]
\label{dfn:discrete_function_space}
Let $X(\grid)$ be a finite-dimensional, piecewise continuous function
space. Let further
\begin{math}
  \left(\basis_E\right)_{E \in \grid}
\end{math}
and
\begin{math}
  \left(\mu_E\right)_{E \in \grid}
\end{math}
be families of local basis function sets and injective local DOF mappings,
\begin{gather*}
  \basis_E = \left\{ \phi_{E,i} \mid i = 0, \ldots, n_E-1 \right\}, \\
  \mu_E: \{0, \ldots, n_E-1\} \rightarrow \{0, \ldots, \dim(X(\grid))-1\},
\end{gather*}
such that
\begin{math}
  X(\grid) = \operatorname{span} \basis_\grid
\end{math}
with the global basis function set $\basis_\grid$ as defined in
\ref{dfn:global_basis}. Then, the triple
\begin{equation*}
  \left(X(\grid), \left(\basis_E\right)_{E \in \grid}, \left(\mu_E\right)_{E \in \grid}\right)
\end{equation*}
is called a \emph{discrete function space}.
\end{definition}

It is important to note that while we motivated the concept of a discrete
function space starting from a discontinuous finite element space
\mbox{Definition \ref{dfn:discrete_function_space}} applies to continuous
finite element spaces as well. For the sake of readability, we will not
always explicitly state the local basis function sets and DOF mappings when
referring to a discrete function space. Instead, we simply speak of a
discrete function space $X(\grid)$ and it is tacitly understood that local
basis function sets
\begin{math}
  \left(\basis_E\right)_{E \in \grid}
\end{math}
and DOF mappings
\begin{math}
  \left(\mu_E\right)_{E \in \grid}
\end{math}
have been fixed.

\begin{example} \label{exp:discrete_function_space}
Let $\bm{k} = (k_E)_{E \in \grid}$ be a vector of local polynomial
degrees. Denote by
\begin{equation*}
  X^{\bm{k}}(\grid) = \left\{ u \in  L^\infty(\Omega) \mid
    u\restr{E} \in P^{k_E}(E) \text{ for all } E \in \grid \right\}
\end{equation*}
the space of piecewise polynomial functions. Assume that for each element
$E \in \grid$ there is a fixed reference element $R_E$ and an affine
reference mapping
\begin{math}
  F_E: R_E \rightarrow E.
\end{math}
We choose
\begin{math}
  \basis_E = \basis_{R_E, k_E} \circ F_E^{-1},
\end{math}
where $\basis_{R, k} \subset C(R)$ denotes the polynomial basis resulting
from a Gram-Schmidt orthonormalization with respect to the $L^2(R)$ inner
product applied to the lexicographically ordered monomials
\begin{math}
  (x^{\bm\alpha})_{\abs{\bm{\alpha}} \leq k},
\end{math}
\begin{equation*}
  x^{\bm\alpha} = \prod_{i = 1}^d x_i^{\alpha_i}
    \quad (\bm\alpha \in \N^d).
\end{equation*}
Note that for the vector space dimensions it holds
\begin{equation*}
  \dim(X^{\bm{k}}(\grid))
    = \sum_{E \in \grid} n_E
    = \sum_{E \in \grid} {k_E + d \choose d}.
\end{equation*}
Assume there is a given strict total order $<$ on $\grid$, e.g., from an
enumeration of the grid elements. Then for each $E \in \grid$ a local DOF
mapping is provided by
\begin{equation*}
  \mu_E(i) = \sum_{\substack{E' \in \grid, \\ E' < E}} n_{E'} + i
    \quad (i = 0, \ldots, n_E-1).
\end{equation*}
\end{example}


\section{Data transfer in \textit{hp}-adaptive computations}
\label{sct:data_transfer}

Adaptive finite element methods give raise to sequences of approximate
solutions, grids, and discrete function spaces. The main challenge we want
to address in this section is the restriction and prolongation of user data
in $hp$-adaptive simulations, i.e., any transfer of discrete functions and
other grid-based data while modifying the mesh or the local ansatz spaces.

In the following, let
\begin{math}
  \left(X^{(m)}\right)_{m \in N}
\end{math}
denote a sequence of discontinuous discrete function spaces,
\begin{equation*}
  X^{(m)} = \left\{ u \in  L^\infty(\Omega) \mid
    u\restr{E} \in X^{(m)}(E) \subset C(E) \text{ for all } E \in \grid^{(m)} \right\}.
\end{equation*}
We assume that the associated sequence of grids
\begin{math}
  \left(\grid^{(m)}\right)_{m \in \N}
\end{math}
is nested in the following sense: for all
\begin{math}
  E^{(m+1)} \in \grid^{(m+1)} \backslash \grid^{(m)}
\end{math}
we assume that either
\vspace{-\topsep}
\begin{enumerate}
  \item there is a unique father element $E^{(m)} \in \grid^{(m)}$ with
  $E^{(m+1)} \subset E^{(m)}$; in this case we say that $E \in
  \grid^{(m+1)}$ resulted from refining $E^{(m)} \in \grid{(m)}$,

  \item or there are a number of elements
  \begin{math}
    E^{(m)}_j \in \grid^{(m)},~j = 1, \ldots, J,
  \end{math}
  such that
  \begin{math}
    E^{(m+1)} = \bigcup_{j = 1}^m E^{(m)}_j;
  \end{math}
  in this case we say that $E \in \grid^{(m+1)}$ resulted from coarsening
  the children $E^{(m)}_j$, $j = 1, \ldots, J$.
\end{enumerate}
\vspace{-\topsep}
Readers familiar with \dune and the definitions in \cite{Bastian2008} may
think of the slightly more general case of sequences of codimension 0 leaf
entity complexes in hierarchical meshes.

For each $m \in \N$ we fix a family of local projection operators
\begin{equation*}
  \Pi^{(m)}_{E}: L^\infty(E) \rightarrow X^{(m)}(E)
    \quad \left(E \in \grid^{(m)}\right).
\end{equation*}
The most important example in view of discontinuous finite element methods
is the following.

\begin{example}[Local $L^2$-projection] \label{exp:local_projection}
For $m \in \N$ and $E \in \grid^{(m)}$ the local $L^2(E)$-projection
\begin{math}
  \Pi^{(m)}_{E}
\end{math}
is defined by
\begin{equation*}
  \int_{E} (\Pi^{(m)}_{E}u) \phi^{(m)} \dd x
    = \int_{E} u \phi^{(m)} \dd x \quad \left(\phi^{(m)} \in X^{(m)}(E)\right).
\end{equation*}
\end{example}

Having fixed a family of local projection operators we denote by
\begin{math}
  \Pi^{(m)}: L^\infty(\Omega) \rightarrow X^{(m)}
\end{math}
the global projection operator defined by
\begin{equation*}
  \Pi^{(m)}u\restr{E} = \Pi^{(m)}_E u\restr{E}
    \quad \left(E \in \grid^{(m)}\right)
\end{equation*}
for all $u \in X^{(m)}$. Now, let
\begin{math}
  u_0 \in L^\infty(\Omega)
\end{math}
be some given initial data. On an abstract level, an adaptive scheme can be
written in the following form,
\begin{align*}
  u^{(0)} & = \Pi^{(0)}u_0, \\
  u^{(m+1)} & = \Pi^{(m+1)} \circ \Phi^{(m)}(u^{(m)}) \quad (m \in \N),
\end{align*}
where each
\begin{math}
  \Phi^{(m)}: X^{(m)} \rightarrow X^{(m)}
\end{math}
is some arbitrary operator. For the remainder of this section we will be
concerned with the restriction and prolongation of a discrete function
$u^{(m)} \in X^{(m)}$, i.e., the efficient computation of the projection
\begin{equation*}
  u^{(m+1)} = \Pi^{(m+1)}u^{(m)}.
\end{equation*}

From a mathematical point of view the restriction and prolongation of a
given function is a trivial task. Its practical implementation, however, is
not, the main difficulty being that during the modification data may be
invalidated. In \dune, the grid adaptation is split into several stages.
During this modification phase user data may be transferred from a grid
state $\grid^{(m)}$ to $\grid^{(m+1)}$. For this particular purpose, a
\dune grid provides persistent id mappings and associative containers, see
\cite{Bastian2008a}. Grid data stored as consecutive arrays (e.g., global
DOF vectors) must be copied into temporary data structures that remain
valid during the adaptation.

The finite element library \dune[Fem] pursues a different strategy. During
the modification phase DOF vectors are resized to hold information
associated with elements $E \in \grid^{(m)} \cup \grid^{(m+1)}$. Each stage
of the adaptation cycle requires updates on index sets, DOF mappings, and
DOF vectors \cite[Algorithms 18sqq.]{Dedner2010}. In case a relatively
small number of elements is marked for local mesh adaptation this approach
leads to significantly less memory overhead during adaptation. A comparison
of the two different adaptation strategies can be found in
\cite{Kloefkorn2009}. Unfortunately, the algorithms and data structures
implemented in \dune[Fem] do not hold in case of $p$- or $hp$-adaptation.
Their generalization must be done with great care if legacy code shall be
supported. For the sake of presentation, we introduce the following
notation.

\begin{definition}[Global DOF set]
For all $m \in \N$ we denote the dimension of the discrete function space
$X^{(m)}$ by
\begin{equation*}
  N^{(m)} = \dim\left(X^{(m)}\right) = \sum_{E \in \grid^{(m)}} n^{(m)}_E,
\end{equation*}
where
\begin{math}
  n^{(m)}_E = \dim X^{(m)}(E).
\end{math}
We define the \emph{global DOF set}
\begin{math}
  \dofs^{(m)} \subset \grid^{(m)} \times \N
\end{math}
associated with $X^{(m)}$ by
\begin{equation*}
  (E,i) \in \dofs^{(m)} :\Leftrightarrow i \in \{0, \ldots, n^{(m)}_E-1 \}.
\end{equation*}
\end{definition}

A \emph{global DOF mapping} is an injective mapping
\begin{math}
  \mu^{(m)}: \dofs^{(m)} \rightarrow \{0, \ldots, N^{(m)}-1 \}.
\end{math}
Obviously, a global DOF mapping $\mu^{(m)}$ is equivalent to a family of
local DOF mappings $\left(\mu^{(m)}_E\right)_{E \in \grid^{(m)}}$ by the
relation
\begin{equation}
  \mu^{(m)}_E(i) = \mu^{(m)}(E,i) \quad \left((E,i) \in \dofs^{(m)}\right).
    \label{eqn:dof_conversion}
\end{equation}
The following definition describes our generalized approach to the
restriction and prolongation of discrete functions.

\begin{definition}[Restriction and prolongation] \label{dfn:restrict_prolong}
Assume a global DOF mapping $\mu^{(m)}$ is already known. Let
\begin{equation*}
  u^{(m)} = \sum_{i = 0}^{N^{(m)}-1} u_i \psi_i^{(m)} \in X^{(m)}
\end{equation*}
be a given discrete function developed in the global basis $\basis_\grid$
defined in \ref{dfn:global_basis}. The global DOF vector will be denoted by
\begin{math}
  \bm{u} = (u_i)_{i = 0, \ldots, N^{(m)}-1},
\end{math}
and we deliberately omit the index $m$. In order to simultaneously compute
a global DOF mapping $\mu^{(m+1)}$ and the projection $u^{(m+1)} =
\Pi^{(m+1)}u^{(m)}$ proceed as follows.

\begin{steps}
\item[Insertion of new DOFs] \label{stp:insert_dofs}
Let
\begin{math}
  \dofs^{(m+1/2)} = \dofs^{(m)} \cup \dofs^{(m+1)}
\end{math}
and $\mu^{(m+1/2)}$ be a continuation of $\mu^{(m)}$ on $\dofs^{(m+1/2)}$,
i.e., $\mu^{(m+1/2)}$ is an injective mapping
\begin{equation*}
  \mu^{(m+1/2)}: \dofs^{(m+1/2)} \rightarrow \{0, \ldots,
  \abs{\dofs^{(m+1/2)}}-1\},
\end{equation*}
such that
\begin{equation*}
  \mu^{(m+1/2)}(E,i) = \mu^{(m)}(E,i)
    \quad \left((E,i) \in \dofs^{(m)}\right).
\end{equation*}

\item[Restriction and prolongation of user data] \label{stp:restrict_prolong}
The global DOF vector $\bm{u}$ is temporarily resized to
$N^{(m+1/2)} = \abs{\dofs^{(m+1/2)}}$. In case of a newly inserted element
\begin{math}
  E \in \grid^{(m+1)} \backslash \grid^{(m)}
\end{math}
created either from local grid refinement or coarsening the associated DOFs
are initialized by
\begin{equation*}
  \sum_{i = 0}^{n^{(m+1)}_E-1} u_{\mu^{(m+1/2)}(E,i)} \phi_{E,i}^{(m+1)}
    = \Pi^{(m+1)} u^{(m)}\restr{E}.
\end{equation*}
If otherwise
$X^{(m+1)}(E) \neq X^{(m)}(E)$ all DOFs
\begin{math}
  u_{\mu^{(m+1/2)}(E,i)},~i = 0, \ldots, n^{(m+1)}_E-1,
\end{math}
must be reinitialized by projecting the local discrete function
$u^{(m)}\restr{E}$ to the local space $X^{(m+1)}(E)$. Note that in general
\begin{math}
  \dofs^{(m)}_E \cap \dofs^{(m+1/2)}_E \neq \emptyset.
\end{math}
We make a copy of the local DOF vector,
\begin{equation*}
  v_i = u_{\mu^{(m+1/2)}(E,i)} \quad (i = 0, \ldots, n^{(m)}_E-1),
\end{equation*}
and compute the projection
\begin{equation*}
  \sum_{i = 0}^{n^{(m+1)}_E-1} u_{\mu^{(m+1/2)}(E,i)} \phi_{E,i}^{(m+1)}
    = \Pi^{(m+1)}_{E} \sum_{i = 0}^{n^{(m)}_E-1} v_{i} \phi^{(m)}_{E,i}.
\end{equation*}

\item[Removal of DOFs] \label{stp:remove_dofs}
Construct a new injective DOF numbering
\begin{equation*}
  \mu^{(m+1)}: \dofs^{(m+1)} \rightarrow \{0, \ldots, N^{(m+1)}-1\},
\end{equation*}
such that
\begin{equation*}
  \mu^{(m+1)}(E,i) = \mu^{(m+1/2)}(E,i)
    \quad \text{if } \mu^{(m+1/2)}(E,i) < N^{(m+1)}.
\end{equation*}
All other DOFs are copied to their new destination,
\begin{equation*}
  u_{\mu^{(m+1)}(E,i)} = u_{\mu^{(m+1/2)}(E,i)}
    \quad \text{if } \mu^{(m+1/2)}(E,i) \geq N^{(m+1)}
\end{equation*}
and the DOF vector $\bm{u}$ is resized to its new length
$N^{(m+1)}$.
\end{steps}
\end{definition}

The algorithm \ref{dfn:restrict_prolong} yields a global DOF mapping
$\mu^{(m+1)}$ and by \mbox{Equation \eqref{eqn:dof_conversion}} the family
of associated local DOF mappings
\begin{math}
  \left(\mu^{(m+1)}_E\right)_{E \in \grid^{(m+1)}}.
\end{math}
At the same time we have computed the projection of $u^{(m)}$ onto
$X^{(m+1)}$,
\begin{equation*}
  u^{(m+1)} = \sum_{i = 0}^{N^{(m+1)}-1} u_i \psi^{(m+1)}_i,
\end{equation*}
developed in the global basis functions defined in \ref{dfn:global_basis}.

Concerning the actual implementation of the above algorithm, \mbox{Steps
\ref{stp:insert_dofs}} and \ref{stp:restrict_prolong} should be clear. The
practical definition of the DOF mapping $\mu^{(m+1)}$ in \mbox{Step
\ref{stp:remove_dofs}}, however, may be in need of further explanation.

\begin{example}[Removal of DOFs]
Let $\mu^{(m+1/2)}$ be the intermediate DOF mapping as in \mbox{Step
\ref{stp:insert_dofs}} above. We denote by
\begin{equation*}
  \mathcal{H}^{(m+1/2)} = \{ 0, \ldots, N^{(m+1)}-1 \}
    \backslash \mu^{(m+1/2)}(\dofs^{(m+1)})
\end{equation*}
the set of freed valid indices, the so-called set of holes. Its elements
$h_i$, $i = 0, \ldots, \abs{\mathcal{H}^{(m+1/2)}}-1$, are assumed to be in
ascending order. Let
\begin{equation*}
  \mu^{(m+1)}: \dofs^{(m+1)} \rightarrow \{ 0, \ldots, \abs{\dofs^{(m+1)}}-1 \}
\end{equation*}
be defined by
\begin{equation*}
  \mu^{(m+1)}(E,j) = \begin{cases}
    \mu^{(m+1/2)}(E,j) & \text{if } \mu^{(m+1/2)}(E,j) < \abs{\dofs^{(m+1)}}-1, \\
    h_i & \text{if } \mu^{(m+1/2)}(E,j) = \abs{\dofs^{(m+1)}} + k_i,
  \end{cases}
\end{equation*}
where the numbers $k_i$, $i = 0, \ldots, \abs{\mathcal{H}^{(m+1/2)}}-1$,
are assumed to be in ascending order as well.
\end{example}

\begin{figure}
  \centering
  \includegraphics[width=0.8\linewidth]{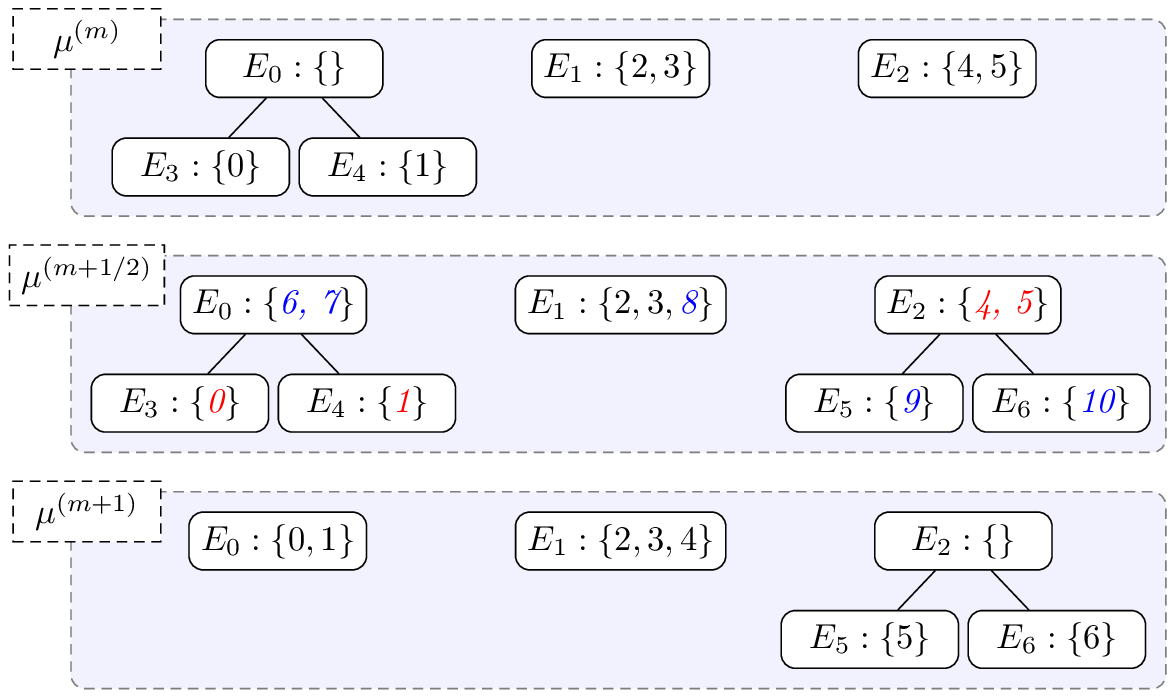}
  \caption{Computation of a global DOF mapping $\mu^{(m+1)}$ by \mbox{Steps
  \ref{stp:insert_dofs}} to \ref{stp:remove_dofs} of \mbox{Definition
  \ref{dfn:restrict_prolong}}, see \mbox{Example
  \ref{exp:restrict_prolong}} for details.}
  \label{fig:restrict_prolong}
\end{figure}

\begin{example} \label{exp:restrict_prolong}
\mbox{Figure \ref{fig:restrict_prolong}} shows a hierarchical grid with
three macro elements $E_0, E_1, E_2$ with $E_0$ having two children $E_3,
E_4$. The leaf level elements form the grid
\begin{math}
  \grid^{(m)} = \{ E_i \mid i = 1, \ldots, 4 \};
\end{math}
the global DOF numbers associated with each element are depicted inside the
brackets. Here, level $0$ elements are assumed to hold at least two DOFs,
while level $1$ elements shall have one DOF. Next, $E_3$ and $E_4$ are
marked for coarsening, $E_2$ is marked for refinement, and $E_1$ gets
assigned an additional DOF by local $p$-refinement. Newly inserted DOFs
added in \mbox{Step \ref{stp:insert_dofs}} and DOFs to be removed in
\mbox{Step \ref{stp:remove_dofs}} are printed in italics. During the
modification phase the number of DOFs is enlarged to allow for the local
projection of user data.
\end{example}

\begin{remark}[Storage costs]
During the adaptation from $X^{(m)}$ to $X^{(m+1)}$ the maximum length of
the global DOF vector equals
\begin{math}
  \abs{\dofs^{(m+1/2)}} = \mathcal{O}(\abs{\grid^{(m)} \cup \grid^{(m+1)}}).
\end{math}
Taking into account the temporary storage used in \mbox{Step
\ref{stp:restrict_prolong}} the overall memory consumption of algorithm
\ref{dfn:restrict_prolong} is of order
\begin{math}
  \mathcal{O}(\abs{\grid^{(m)} \cup \grid^{(m+1)}} + 1).
\end{math}
\end{remark}


\section{Using the dune-fem-hpdg module}
\label{sct:dune_fem_hpdg}

In this section we describe the usage of the \texttt{dune-fem-hpdg} add-on
module to the \dune[Fem] finite element library. The module provides
extensible reference implementations of adaptive discrete function spaces
for implementing $p$- and $hp$-adaptive discontinuous finite element
methods. First, we give a brief introduction to \dune[Fem] and in
particular its local mesh adaptation capabilities.

\subsection{The Dune-Fem finite element library}
\label{sct:dune_fem}

\begin{figure}
  \centering
  \includegraphics[width=0.8\linewidth]{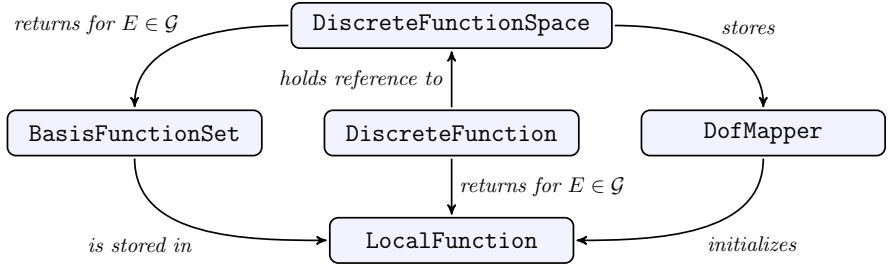}
  \caption{Interplay of the main interface classes in \dune[Fem]
  representing discrete function spaces and discrete functions.}
  \label{fig:interface_classes}
\end{figure}

The \Cpp-library \dune[Fem] provides a number of abstract interface classes
representing discrete functions and discrete function spaces, local basis
function sets, and DOF vectors and DOF mappings. These classes and their
relationships are summarized in \mbox{Figure \ref{fig:interface_classes}}.
Here, a \lstinline!DiscreteFunction! represents an element
\begin{math}
  u \in X(\grid).
\end{math}
A discrete function holds a global DOF vector
\begin{equation*}
  \bm{u} = (u_0, \ldots, u_{N-1})
\end{equation*}
and is usually evaluated in local coordinates with respect to a given
element $E \in \grid$. The local DOF vector
\begin{equation*}
  (u_{\mu_E(0)}, \ldots, u_{\mu_E(n_E-1)})
\end{equation*}
is initialized by the \lstinline!DofMapper!, a class representing the family of
local DOF mappings
\begin{math}
  \left( \mu_E \right)_{E \in \grid}.
\end{math}
Finally, a \lstinline!BasisFunctionSet! represents the local basis function set
$\basis_E$, and the local function $u\restr{E} \in X(E)$ is given by
\begin{equation*}
  u\restr{E} = \sum_{i = 0}^{n_E-1} u_{\mu_E(i)} \phi_{E,i}.
\end{equation*}

In most complex applications, more than one discrete function is in use.
This poses some difficulty in adaptive simulations. A grid may be only be
adapted once, and all discrete functions must be restricted and prolonged
simultaneously. In \dune[Fem], a sole class, the \lstinline!DofManager!, is
responsible for starting the grid adaptation process. It resizes all DOF
vectors and notifies DOF mappings about the update.

In case no user data must be restricted and prolonged no action is
required. Otherwise, users must add discrete functions, containers, etc.,
to an \lstinline!AdaptationManager!. This class is equipped with the
necessary information (encapsulated in a \lstinline!RestrictProlong!
object) on how to locally project user data during adaptation.
\mbox{Listing \ref{lst:h_adaptation}} illustrates the adaptation process
from a user perspective.

\begin{lstlisting}[
  float, style=float,
  caption={Sample code illustrating the automated restriction and
  prolongation of discrete functions under grid refinement in
  \textsc{Dune-Fem}.},
  label={lst:h_adaptation},
]
template <class DiscreteFunction>
void adapt(DiscreteFunction &uh, unsigned int seed) {
  // get grid (slightly simplified)
  using Grid = typename DiscreteFunction::GridType;
  Grid &grid = uh.grid();

  // create adaptation manager (slightly simplified)
  using RestrictProlong = RestrictProlongDefault<DiscreteFunction>;
  using h_AdaptationManager = AdaptationManager<Grid, RestrictProlong>;
  h_AdaptationManager h_adaptManager(RestrictProlong(uh));

  // randomly mark leaf level grid elements for refinement/coarsening
  default_random_engine engine(seed);
  uniform_int_distribution<int> distribution(-1, 1);
  for (const auto &element : elements(grid.leafGridView()))
    grid.mark(distribution(engine), element);

  // adapt discrete function
  h_adaptManager.adapt();
}
\end{lstlisting}

\subsection{Discrete function spaces in dune-fem-hpdg}
\label{sct:dune_fem_hpdg_spaces}

As illustrated in \mbox{Figure \ref{fig:interface_classes}} a
\lstinline!DiscreteFunctionSpace! assembles the local basis function sets and DOF
mappings. The \texttt{dune-fem-hpdg} extends the list of available discrete
function spaces available to users of \dune[Fem] by the following
$hp$-adaptive spaces:
\vspace{-\topsep}
\begin{description}[labelindent=1em, font=\texttt]
  \item[OrthogonalDiscontinuousGalerkinSpace] an implementation of the
  example finite element space \ref{exp:discrete_function_space},

  \item[LegendreDiscontinuousGalerkinSpace] a discontinuous finite element
  space using product Legendre ansatz polynomials, i.e.,
  \begin{math}
    \basis_{E,k} = \left\{ \phi_{\bm{\alpha}}
      \mid 0 \leq \alpha_i \leq k, i = 1, \ldots, d \right\}, k \in \N,
  \end{math}
  with
  \begin{equation*}
    \phi_{\bm{\alpha}}(F_E(x)) = \prod_{i = 1}^d p_{\alpha_i}(x_i)
      \quad (x \in R_E),
  \end{equation*}
  where $F_E: R_E \rightarrow E$ is an affine reference mapping and the
  $i$-th Legendre polynomial on the unit interval $[0,1]$ is defined by
  \begin{equation*}
    p_i(x) = \frac{1}{i!}
      \frac{\mathrm{d}^i}{\mathrm{d}x^i}\bigl[(x^2 - x)^i\bigr],
  \end{equation*}

  \item[AnisotropicDiscontinuousGalerkinSpace] a discontinuous space based
  on product Legendre polynomials as well; this implementation, however,
  allows for an adaptation in $p$ in each spatial direction.
\end{description}
\vspace{-\topsep}
We remark that the latter two spaces based on Legendre polynomials are
restricted to cubic grids, i.e., $R_E = [0,1]^d$ for all $E \in \grid$.

The discrete function spaces listed above differ in the choice of local
basis function sets. They do, however, share the common base class
\lstinline!hpDG::DiscontinuousGalerkinSpace! providing almost all other
functionality. In the same way, the \texttt{dune-fem-hpdg} module allows users
to quickly setup new discrete function spaces. In order to do so, users
must provide an implementation of the \lstinline!hpDG::BasisFunctionSets!
interface which represents a family of local basis function sets. This is
explained in more detail in \mbox{Appendix \ref{apx:user_api}}.

The construction of the discontinuous finite element spaces in
\texttt{dune-fem-hpdg} is in no way different from that of all other
discrete function spaces in \dune[Fem]. \mbox{Listing
\ref{lst:construction}} illustrates the initialization of an adaptive
discrete function suitable for $hp$-adaptive simulations.

\begin{lstlisting}[
  float, style=float,
  caption={Construction of a discrete function space and a discrete
  function with \texttt{dune-fem-hpdg} suitable for $hp$-adaptive
  simulations.},
  label={lst:construction}
]
// create quadrilateral ALUGrid
using Grid = ALUGrid<2, cube, nonconforming>;
Grid grid(/*...*/);

// select leaf level partition
using GridPart = AdaptiveLeafGridPart<Grid>;
GridPart gridPart(grid);

// create scalar discrete function space
const int order = 3;
using DiscreteFunctionSpace = hpDG::OrthogonalDiscontinuousGalerkinSpace<
    FunctionSpace<double, double 2, 1>, GridPart, order>;
DiscreteFunctionSpace space(gridPart);

// create discrete function
using DiscreteFunction = AdaptiveDiscreteFunction<DiscreteFunctionSpace>;
DiscreteFunction uh("uh", space);
\end{lstlisting}

\subsection{An interface for local \textit{p}-adaptation}
\label{sct:dune_fem_hpdg_p_adaption}

\begin{lstlisting}[
  float, style=float,
  caption={Sample code illustrating the automated restriction and
  prolongation of discrete functions under modification of the local
  polynomial degree.},
  label={lst:p_adaptation}
]
template <class DiscreteFunction>
void adapt(DiscreteFunction &uh, unsigned int seed) {
  // get space (slightly simplified)
  using DiscreteFunctionSpace =
      typename DiscreteFunction::DiscreteFunctionSpaceType;
  DiscreteFunctionSpace &space = uh.space();

  // create adaptation manager
  using Grid = typename DiscreteFunction::GridType;
  using DataProjection = hpDG::DefaultDataProjection<DiscreteFunction>;
  using p_AdaptationManager = hpDG::AdaptationManager<Grid, DataProjection>;
  p_AdaptationManager p_adaptManager(DataProjection(uh));

  // randomly mark grid (part) elements for p-adaptation
  default_random_engine engine(seed);
  uniform_int_distribution<int> distribution(0, space.order());
  for (const auto &element : elements(uh.gridPart()))
    space.mark(distribution(engine), element);

  // adapt discrete function
  p_adaptManager.adapt();
}
\end{lstlisting}

The most important design decision we made in implementing the
\texttt{dune-fem-hpdg} module was to split $h$- and $p$-refinement in two
separate stages. Our reasoning is twofold. First, the software should
support $h$-adaptive legacy code in the manner described above in
\mbox{Section \ref{sct:dune_fem}}. Second, this strategy allows for an
arbitrary number of $h$-, $p$- and $hp$-adaptive discrete functions in
single application and gives users the necessary freedom in complex
applications. Standard use cases are easy to implement as will be shown in
\mbox{Section \ref{sct:numerical_results}}.

Any of the discrete function spaces in \texttt{dune-fem-hpdg} provides a set
of extended interface methods for local $p$-adaptation. In most papers, $p$
denotes an integer, e.g., the polynomial order of a local approximation
space. We slightly generalized this idea and allow $p$ to be of arbitrary
type (the \lstinline!Key! type), e.g., a vector of individual local polynomial
degrees for each space direction. The $p$-adaptive interface mimicks that
of the grid adaptation in \dune.
\begin{lstlisting}[style=display]
const DiscreteFunctionSpace::KeyType &key(
    const DiscreteFunctionType::EntityType &entity) const;
void DiscreteFunctionSpace::mark(
    const DiscreteFunctionType::KeyType &key,
    const DiscreteFunctionType::EntityType &element);
\end{lstlisting}
The first method returns the key currently assigned to a grid element,
while \lstinline!mark! allows for re-assigning a new key. In order for the
changes to take effect, one of the following two \lstinline!adapt! methods must
be called.
\begin{lstlisting}[style=display]
bool DiscreteFunctionSpace::adapt();
template <DataProjection>
bool DiscreteFunctionSpace::adapt(DataProjection &projection);
\end{lstlisting}
The necessary information on how to locally project data are encapsulated
in a \lstinline!DataProjection!. As in case of $h$-adaptation, however,
users will not call these methods explicitly. Instead, the
\lstinline!hpDG::AdaptationManager! class handles the restriction and
prolongation as illustrated by \mbox{Listing \ref{lst:p_adaptation}}. Note
how closely the code resembles \mbox{Listing \ref{lst:h_adaptation}} above
on local mesh adaptation in \dune[Fem]. In this particular example, a
single discrete function is restricted and prolonged; however, the
\texttt{dune-fem-hpdg} facilities may be provided with an arbitrary number
of discrete functions and other user data. More information on the
adaptation and the restriction and prolongation of custom user data can be
found in \mbox{Appendix \ref{apx:user_api}}.


\section{An \textit{hp}-adaptive interior penalty Galerkin method}
\label{sct:numerical_results}

In this final section we want to illustrate in a complex application the
capabilities and usage of the \texttt{dune-fem-hpdg} module. We will be
concerned with the numerical solution of an elliptic PDE by means of an
$hp$-adaptive symmetric interior penalty Galerkin (SIPG) scheme. First, we
briefly revisit the $hp$-version of the SIPG method following the recent
book by Dolej\v{s}\'{\i} and Feistauer \cite[Chapter 7]{Dolejsi2015}.

\subsection{The \textit{hp}-version of the SIPG method}

Let $\Omega \subset \R^2$ be a bounded domain with Lipschitz boundary
$\partial \Omega$. We consider the following elliptic model problem
\begin{equation}
\begin{aligned}
  -\Delta u & = f \quad \text{in } \Omega,
          u & = g \quad \text{on } \partial \Omega
\end{aligned}
\label{eqn:poisson}
\end{equation}
for given source $f$ and boundary values $g$. The domain is discretized by
a grid $\grid$; for $E \in \grid$ we denote the sets of interior and
boundary intersections by
\begin{align*}
  \intersections[int]{E}
    & = \{ e = E \cap E' \mid E' \in \grid\backslash\{ E \} \}, \\
  \intersections[bnd]{E}
    & = \{ e = \partial E \cap \partial \Omega \}.
\end{align*}
The sets of all interior and boundary intersections will be denoted by
$\Gamma_{\mathit{int}}$ and $\Gamma_{\mathit{bnd}}$, respectively. We fix
local polynomial degrees
\begin{math}
  \bm{k} = (k_E)_{E \in \grid}
\end{math}
with $k_E \geq 1$ for all $E \in \grid$ and consider the standard
discontinuous finite element space
\begin{equation*}
  X^{\bm{k}}(\grid)
    = \{ u \in L^\infty(\Omega) \mid u\restr{E} \in P^{k_E}(E)
      \text{ for all } E \in \grid \}.
\end{equation*}
The jump of a discrete function
\begin{math}
  u \in X^{\bm{k}}(\grid)
\end{math}
across an inter-element intersection $e \in \Gamma_{\mathit{int}}$, $e
= E \cap E'$, is defined by
\begin{equation*}
  \llbracket u \rrbracket(x) = \left(u\restr{E} \nu_E
    + u\restr{E'} \nu_{E'}\right)(x) \quad (x \in e),
\end{equation*}
where $\nu_E$ denotes a unit outer normal to $E \in \grid$. The average of
a discrete function is given by
\begin{equation*}
  \{ u \}(x) = \frac{1}{2} \bigl(u\restr{E} + u\restr{E'}\bigr)(x)
  \quad (x \in e).
\end{equation*}
For each intersection
\begin{math}
  e \in \Gamma_{\mathit{int}} \cup \Gamma_{\mathit{bnd}}
\end{math}
we introduce the penalty parameter
\begin{equation*}
  \sigma_e
    = \begin{dcases}
        \gamma \frac{k_E^2 + k_{E'}^2}{2 h_e} & \text{for } e = E \cap E', \\
        \gamma \frac{k_E^2}{h_e} & \text{for } e = \partial E \cap \partial \Omega,
      \end{dcases}
\end{equation*}
where $\gamma > 0$ is a sufficiently large constant and $h_e$ denotes the
diameter of $e$. Note that the penalty parameter depends on the local
polynomial degrees. Now, let
\begin{math}
  B: X^{\bm{k}}(\grid) \times X^{\bm{k}}(\grid) \rightarrow \R
\end{math}
be a bilinear form defined by
\begin{equation*}
  \begin{split}
    B(u, \psi)
        = & \sum_{E \in \grid} \int_{E} \nabla u \cdot \nabla \psi \dd x \\
      & \quad - \sum_{e \in \Gamma_{\mathit{int}}} \int_{e} \jump{u}
      \mean{\nabla \psi} + \jump{\psi} \mean{\nabla u} \dd s
              + \sum_{e \in \Gamma_{\mathit{int}}} \int_{e} \sigma_e
              \jump{u} \jump{\psi} \dd s \\
      & \quad - \sum_{e \in \Gamma_{\mathit{bnd}}} \int_{e} u (\nabla \psi
      \cdot \nu) + \psi (\nabla u \cdot \nu) \dd s
              + \sum_{e \in \Gamma_{\mathit{bnd}}} \int_{e} \sigma_e u \psi
              \dd s,
  \end{split}
\end{equation*}
and
\begin{math}
  l: X^{\bm{k}}(\grid) \rightarrow \R
\end{math}
defined by
\begin{equation*}
  l(\psi)
    = \int_{\Omega} f \psi \dd x
        - \sum_{e \in \Gamma_{\mathit{bnd}}} \int_{e} g (\nabla \psi \cdot
        \nu) \dd s
        + \sum_{e \in \Gamma_{\mathit{bnd}}} \int_{e} \sigma_e g \psi \dd s.
\end{equation*}
The bilinear form $B$ is continuous and coercive with respect to the energy
norm
\begin{math}
  \norm{\cdot}_{\mathit{DG}}: X^{\bm{k}}(\grid) \rightarrow \R
\end{math}
defined by
\begin{equation*}
    \norm{u}_{\mathit{DG}}^2
      = \sum_{E \in \grid} \int_E \abs{\nabla u}^2 \dd x
          + \sum_{e \in \Gamma_{\mathit{int}}} \int_{e} \sigma_e \jump{u}^2
          \dd s
          + \sum_{e \in \Gamma_{\mathit{bnd}}} \int_{e} \sigma_e u^2 \dd s
\end{equation*}
provided the constant $\gamma$ is sufficiently large \cite[Theorems 7.13
and 7.15]{Dolejsi2015}. This guarantees the existence and uniqueness of the
weak solution $u_h \in X^{\bm{k}}(\grid)$ to the model problem
\eqref{eqn:poisson} defined by
\begin{equation}
  B(u_h, \psi) = l(\psi) \quad (\psi \in X^{\bm{k}}(\grid)).
    \label{eqn:hp_sipg_numerical_solution}
\end{equation}

\subsection{The \textit{hp}-adaptive scheme}

The first component in the construction of a fully $hp$-adaptive scheme is
an \emph{a posteriori} error indicator estimating the local approximation
error. Of course, the local error indicators must be computable from the
numerical solution $u_h$ and the given data $f, g$ alone. There is a
growing body of literature on \emph{a posteriori} error estimation, see,
e.g., \cite{Dolejsi2013, Houston2007, Zhu2011}. We implemented the
following indicator from \cite{Houston2008},
\begin{equation}
\begin{split}
  \eta_E^2
    = & \frac{{h_E}^2}{{k_E}^2} \int_E \bigl(\Pi_{E, k_E-1}(f + \Delta
    u_h)\bigr)^2 \dd x \\
      & \quad + \sum_{e \in \intersections[\mathit{int}]{E}}
          \left\{\frac{h_E}{k_E} \int_e \bigl(\Pi_{e, k_e-1} \jump{\nabla
          u_h \cdot \nu_E}\bigr)^2 \dd s
            + \frac{{k_E}^3}{h_E} \int_e \jump{u_h}^2 \dd s \right\} \\
      & \quad + \sum_{e \in \intersections[\mathit{bnd}]{E}}
      \frac{{k_E}^3}{h_E} \int_e (u_h - g)^2 \dd s \quad (E \in \grid).
\end{split}
\label{eqn:hp_sipg_eta}
\end{equation}
Here,
\begin{math}
  \Pi_{E, k_E-1} \text{ and } \Pi_{e, k_e-1}
\end{math}
denote local $L^2$-projections onto the polynomial spaces of lower degree
$P^{k_E - 1}(E)$ and $P^{k_e - 1}(e)$ with $k_e = \max\{k_E, k_{E'}\}$. It
can be shown that
\begin{equation*}
  \norm{u - u_h}_{\mathit{DG}}
    \leq \left( \sum_{E \in \grid} \eta_E^2 + \mathcal{O}(f, u_h) \right)^{\frac{1}{2}},
\end{equation*}
where
\begin{math}
  \mathcal{O}(f, u_h)
\end{math}
denotes a data-oscillation term, see \cite[Theorem 3.2]{Houston2008} for
details. Having computed the local error indicator we mark $E \in \grid$
for refinement in either $h$ or $p$ if the local error indicator exceeds an
upper bound $\eta^*$,
\begin{equation*}
  \eta_E < \eta_* = \mathit{TOL}/\abs{\grid}.
\end{equation*}
If no element is marked for further refinement we stop the iterative
procedure.

Once a grid element has been identified for local adaptation it must be
decided whether to refine in $h$ or $p$. Several strategies for this have
been proposed in the literature, see, e.g., \cite{Demkowicz2002,
Eibner2007, Houston2003, Houston2005, Mitchell2014, Solin2004}. We
implemented the so-called \textsc{Prior2P} strategy described in
\cite{Mitchell2011} based on the following idea. \emph{A priori} error
estimates (see, e.g., \cite[Theorem 7.20]{Dolejsi2015}) suggest that we
should increase the local polynomial degree $k_E$ provided the exact is
sufficiently smooth in $E \in \grid$. A regularity indicator yields an
estimate for the local Sobolev index $q_E$,
\begin{equation*}
  q_E = \max \{ q \mid u\restr{E} \in H^{q}(E) \} \quad (E \in \grid).
\end{equation*}
We increase the local polynomial order provided $k_E < q_E-1$; otherwise,
we mark $E$ for local mesh refinement.

\begin{algorithm}[t]
\begin{algorithmic}[1]
\State choose initial grid $\grid$ and local polynomial degrees
  $\bm{k} = \left(k_E\right)_{E \in \grid}$
\For{$m = 0, 1, \ldots$}
  \State solve system \eqref{eqn:hp_sipg_numerical_solution} to compute
    $u_h \in X^{\bm{k}}(\grid)$
  \ForAll{elements $E \in \grid$}
    \State compute the local error indicator $\eta_E$ from \eqref{eqn:hp_sipg_eta}
  \EndFor
  \If{$\left(\sum_{E \in \grid} \eta_E^2\right)^{\frac{1}{2}} \leq \mathit{TOL}$}
    \State \textbf{stop}
  \EndIf
  \ForAll{elements $E \in \grid$} \label{alg:hp_sipg_1}
    \If{$\eta_E < \eta_* $}
       \State mark element $E$ for $h$-coarsening, if possible;
       \State otherwise, set $k_E \gets \max\{ k_E-1, k_{\textit{min}} \}$
    \ElsIf{$\eta_E > \eta^* $}
       \State compute estimate for the local Sobolev index $q_E$
       \If{$q_E > k_E+1$}
         \State set $k_E \gets \min\{ k_E+1, k_{\textit{max}} \}$
       \Else
         \State mark element $E$ for $h$-refinement
      \EndIf
    \EndIf
  \EndFor \label{alg:hp_sipg_2}
  \While{adapting the grid $\grid$}
    \State restrict and prolong $u_h$ and $\bm{k}$ according to
    \eqref{eqn:sipg_k_prolongation} and \eqref{eqn:sipg_k_restriction}
  \EndWhile
\EndFor
\end{algorithmic}
\caption{Summary of the $hp$-adaptive SIPG method.}
\label{alg:hp_sipg}
\end{algorithm}

So far we have only discussed local $hp$-refinement. Assume now that for $E
\in \grid$ the estimated local approximation error is small, i.e.,
\begin{equation*}
  \eta_E \leq \eta_* \ll \eta^*.
\end{equation*}
We may decide to reduce the number of degrees of freedom by decreasing the
local polynomial degree $k_E$ or by coarsening the grid. Mesh coarsening
usually depends on a number of neighboring elements (e.g., the children of
a common father element in a hierarchical grid) all being marked for grid
coarsening. We take a rather hands-on approach and --- if possible ---
always favor $h$- over $p$-coarsening. We remark that the regularity
indicator implemented restricts the minimum local polynomial degree to
$k_{\mathit{min}} = 3$.

Finally, we need to address the restriction and prolongation of the vector
of local polynomial degrees $\bm{k}$. Let $E \in \grid^{(m+1)}
\backslash \grid^{(m)}$ be an element created during grid modification. We
must define the local polynomial degree $k^{(m+1)}_E$ to be associated with
the newly created element. In case $E$ results from local mesh refinement
of $E' \in \grid^{(m)}$ we simply set
\begin{equation}
  k^{(m+1)}_{E} = k^{(m)}_{E'}
  \label{eqn:sipg_k_prolongation}
\end{equation}
locally prolonging $\bm{k}^{(m)}$ to $\bm{k}^{(m+1)}$. If
otherwise $E$ results from local mesh coarsening we locally restrict
$\bm{k}^{(m)}$ to $\bm{k}^{(m+1)}$ by setting
\begin{equation}
  k^{(m+1)}_{E}
    = \max_{\substack{E' \in \grid^{(m)}, \\ E' \subset E}} k^{(m)}_{E'}.
  \label{eqn:sipg_k_restriction}
\end{equation}
The complete $hp$-adaptive SIPG scheme is summarized in \mbox{Algorithm
\ref{alg:hp_sipg}}.

\subsection{Implementation details}

The \texttt{dune-fem-hpdg} contains a reference implementation of the
$hp$-adaptive SIPG scheme described above. For the computation of the
numerical solution we relied on \dune and the \dune[Fem] discretization
module. The latter provides sample implementations of continuous and
discontinuous finite element methods including an SIPG method for fixed
global polynomial degree $k$. For the solution of the linear system
\eqref{eqn:hp_sipg_numerical_solution} we used the \underline{I}terative
\underline{S}olver \underline{T}emplate \underline{L}ibrary (\dune[ISTL]),
a \dune module developed by Blatt and Bastian \cite{Blatt2007, Blatt2008}.
The $hp$-adaptive scheme requires for a locally adaptive grid. In all our
experiments we used the \dune[ALUGrid] module by Alk{\"a}mper et al.
\cite{Alkaemper2016}, a \dune add-on encapsulating the ALUGrid library by
Schupp \cite{Schupp1999}.

We want to give some details on the implementation in order to illustrate
the $hp$-adaptation process. \mbox{Listing \ref{lst:hp_marking}} shows a
sample implementation of the marking procedure (Lines \ref{alg:hp_sipg_1}
to \ref{alg:hp_sipg_2} of \mbox{Algorithm \ref{alg:hp_sipg}}).

\begin{lstlisting}[
  float, style=float,
  caption={Implementation of Algorithm \ref{alg:hp_sipg}, Lines
  \ref{alg:hp_sipg_1} to \ref{alg:hp_sipg_2}. Note that the given vector of
  polynomial degrees $k$ associated with the discontinuous finite element
  space $X^{(m)}$ is overridden.},
  label={lst:hp_marking}
]
enum class Flag { Coarsen, None, Refine };

template <class Grid, class ManagedArray, class Function>
void mark(Grid &grid, ManagedArray &k, Function function) {
  for (const auto &element : elements(grid.leafGridView())) {
    // get refinement flag and estimated Sobolev index
    Flag what;
    int q;
    tie(what, q) = function(element);

    // mark grid/vector of local polynomial degrees for
    // hp-refinement/coarsening
    switch (what) {
      case Flag::None:
        break;
      case Flag::Coarsen:
        if (element.hasFather())
          grid.mark(-1, element);
        else if (k[element] > 3)
          k[element] -= 1;
        break;
      case Flag::Refine:
        if (k[element] < q - 1)
          k[element] += 1;
        else
          grid.mark(1, element);
    }
  }
}
\end{lstlisting}

Here, the parameter \lstinline!function! is a callable object which returns
a pair of a refinement flag and an estimate for the local Sobolev index for
a given entity. The array \lstinline!k! is an associative container of
local polynomial degrees currently in use. In case an element is flagged
for refinement or coarsening it is either marked for $h$- or
$p$-adaptation; in the latter case we re-assign its local polynomial degree
in \lstinline!k!. The modification of the grid and the finite element space
as well as the restriction and prolongation of the discrete data are shown
in \mbox{Listing \ref{lst:hp_adapt}}.

\begin{lstlisting}[
  float, style=float,
  caption={Prolongation of the vector of polynomial degrees during grid
  adaptation. It is assumed that the function \texttt{mark} from Listing
  \ref{lst:hp_marking} has been called in advance.},
  label={lst:hp_adapt}
]
template <class DiscreteFunction, class ManagedArray>
void adapt(DiscreteFunction &u, ManagedArray &k) {
  // get discrete function space (slightly simplified)
  using DiscreteFunctionSpace =
      typename DiscreteFunction::DiscreteFunctionSpaceType;
  DiscreteFunctionSpace &space = u.space();

  // get hierarchical grid
  using Grid = typename DiscreteFunctionSpace::GridType;
  Grid &grid = space.grid();

  // h-adaptation, restrict/prolong numerical solution and polynomial degrees
  using RestrictProlong =
      RestrictProlongDefaultTuple<DiscreteFunction, ManagedArray>;
  using h_AdaptationManager = AdaptationManager<Grid, RestrictProlong>;
  h_AdaptationManager h_AdaptManager(grid, RestrictProlong(u, k));
  h_AdaptManager.adapt();

  // equip discrete function space with new local polynomial degrees
  for (const auto &element : elements(grid.leafGridView()))
    space.mark(k[element], element);

  // p-adaptation, restrict/prolong numerical solution
  using DataProjection = hpDG::DefaultDataProjection<DiscreteFunction>;
  using p_AdaptationManager =
      hpDG::AdaptationManager<DiscreteFunctionSpace, DataProjection>;
  p_AdaptationManager p_AdaptManager(space, DataProjection(u));
  p_AdaptManager.adapt();
}
\end{lstlisting}

The first input argument of \lstinline!adapt! is the approximate solution
\begin{math}
  u_h^{(m)} \in X^{(m)} = X^{\bm{k}^{(m)}}(\grid^{(m)}).
\end{math}
First, the grid is adapted to its new state $\grid^{(m+1)}$. The
approximate solution and the container are restricted
$\bm{k}^{(m)}$ by the usual \dune[Fem] facilities. This yields the
vector $\bm{k}^{(m+1)}$ and an intermediate projection
\begin{math}
  u_h^{(m+1/2)} \in X^{\bm{k}^{(m)}}(\grid^{(m+1)}).
\end{math}
In Lines 20 and 21 the finite element space is marked for the concluding
$p$-adaptation. The subsequent restriction and prolongation is handled by
\texttt{dune-fem-hpdg}. It yields the desired projection of $u_h^{(m)}$ onto
$X^{(m+1)}$. The resulting function serves as initial guess in the solution
of the next linear system \eqref{eqn:hp_sipg_numerical_solution}.

\subsection{Numerical results}

We consider the homogeneous model problem to recover a prescribed solution
given by
\begin{equation*}
  u(x, y) = r^{2/3}\sin(2\varphi/3)
\end{equation*}
in a domain with reentrant corner
\begin{math}
  \Omega = (-1,1)^2 \backslash (0,-1) \times (1,0).
\end{math}
We compute two series of numerical solutions $u^{(m)}_h$, $m = 0, \ldots,
8$, on an axis-aligned quadrilateral mesh and a triangular mesh. In each
case the $hp$-adaptive SIPG scheme yields a sequence of locally adapted
meshes and distributions of local polynomial degrees. The so-called
$hp$-meshes are shown in \mbox{Figures
\ref{fig:hp_sipg_quadrilateral_1},\ref{fig:hp_sipg_quadrilateral_2}} and
\ref{fig:hp_sipg_triangular_1},\ref{fig:hp_sipg_triangular_2}. Since the
exact solution is known we can compute the approximation error
\begin{equation*}
  e_h^{(m)} = u - u_h^{(m)}.
\end{equation*}
In order to determine the convergence rate (EOC) we follow
\cite{Dolejsi2015} and define
\begin{equation*}
  \mathrm{EOC} = - \frac{\log(\norm{e_h^{(m)}}/\norm{e_h^{(m+1)}})}{\log((N^{(m)}/N^{(m+1)})^{1/d})},
\end{equation*}
where $N^{(m)}$ denotes the number of global DOFs. Another interesting
quantity for evaluating the effectiveness of the adaptive scheme is the
so-called effectivity index, defined as the ratio of the \emph{a
posteriori} error indicator error bound and the energy norm, i.e.,
\begin{equation*}
  \text{Eff. index} = \frac{\left(\sum_{E \in \grid^{(m)}}
    \eta_E^{(m)^2}\right)^{\frac{1}{2}}}{\norm{e_h^{(m)}}_{\mathit{DG}}}.
\end{equation*}
The effectivity index depends on the problem under consideration and the
macro grid. Within the elements of given series of approximate solutions,
however, the effectivity index should be a constant. The approximation
errors, convergence rates, and effectivity indices for the model problem
are shown in \mbox{Tables \ref{tbl:hp_sipg_quadrilateral}} and
\ref{tbl:hp_sipg_triangular}. We encourage users to run the sample code
themselves; the instructions can be found in \mbox{Appendix
\ref{apx:instructions}}.


\appendix

\section{Installing and running the software}
\label{apx:instructions}

In this appendix we describe how to download and install the
\texttt{dune-fem-hpdg} module and how to reproduce the numerical results
presented in \mbox{Section \ref{sct:numerical_results}}.

\subsection{Download of required Dune modules}

The following \dune modules are needed in order to build and run the
examples:
\vspace{-\topsep}
\begin{enumerate}
  \item the \dune 2.4 core modules \dune[Common], \dune[Geometry],
  \dune[Grid], \dune[ISTL], and
  \dune[LocalFunctions]\footnote{\url{http://www.dune-project.org/download.html}},

  \item the \dune[ALUGrid]
  library\footnote{\url{https://gitlab.dune-project.org/extensions/dune-alugrid}},

  \item the \dune[Fem] discretization
  module\footnote{\url{https://gitlab.dune-project.org/dune-fem/dune-fem}},

  \item and the \texttt{dune-fem-hpdg}
  module\footnote{\url{https://gitlab.dune-project.org/christoph.gersbacher/dune-fem-hpdg.git}}.
\end{enumerate}
\vspace{-\topsep}
Please make sure that you check out the \dune 2.4 compatible (release)
branches of \dune[ALUGrid], \dune[Fem], and \texttt{dune-fem-hpdg}.

\subsection{Installation of the software}

Before installing \dune and its components please refer to the installation
notes\footnote{\url{http://www.dune-project.org/doc/installation-notes.html}}
for a list of dependencies and required software. We assume all
aforementioned \textsc{Dune} packages have been downloaded and saved to a
single directory \texttt{\$DUNE}. For the simultaneous build of all modules
using CMake run the \texttt{dunecontrol} script, e.g., by
\begin{lstlisting}[style=display]
cd $DUNE
./dune-common/bin/dunecontrol all
\end{lstlisting}
Please refer to the installation notes for more information and on how to
pass options to the build process.

\subsection{Running the \textit{hp}-adaptive sample code}

\begin{figure}[p]
  \centering
  \begin{minipage}{0.8\linewidth}
    \centering
    \includegraphics[width=0.49\linewidth]{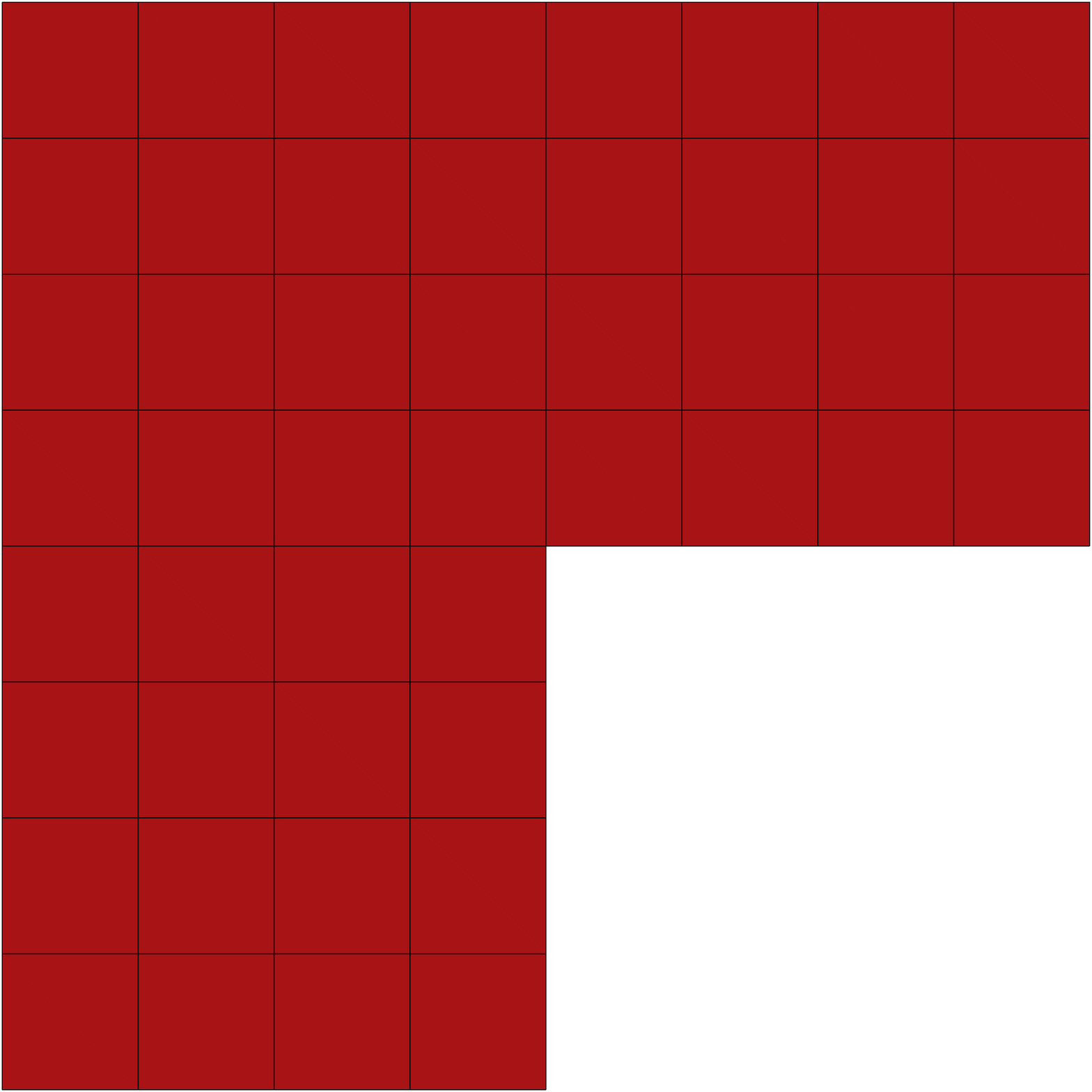}
    \includegraphics[width=0.49\linewidth]{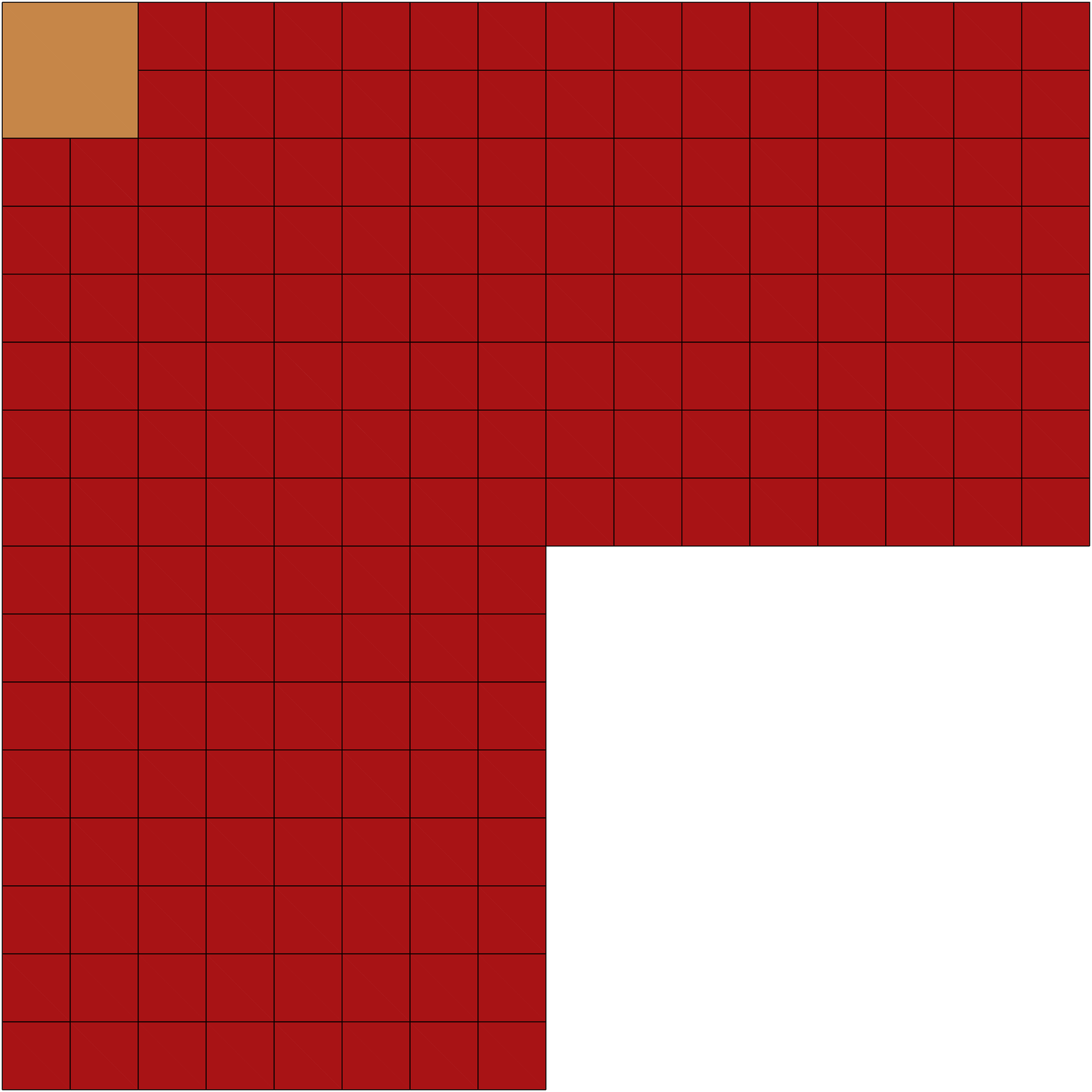} \\
    \includegraphics[width=0.49\linewidth]{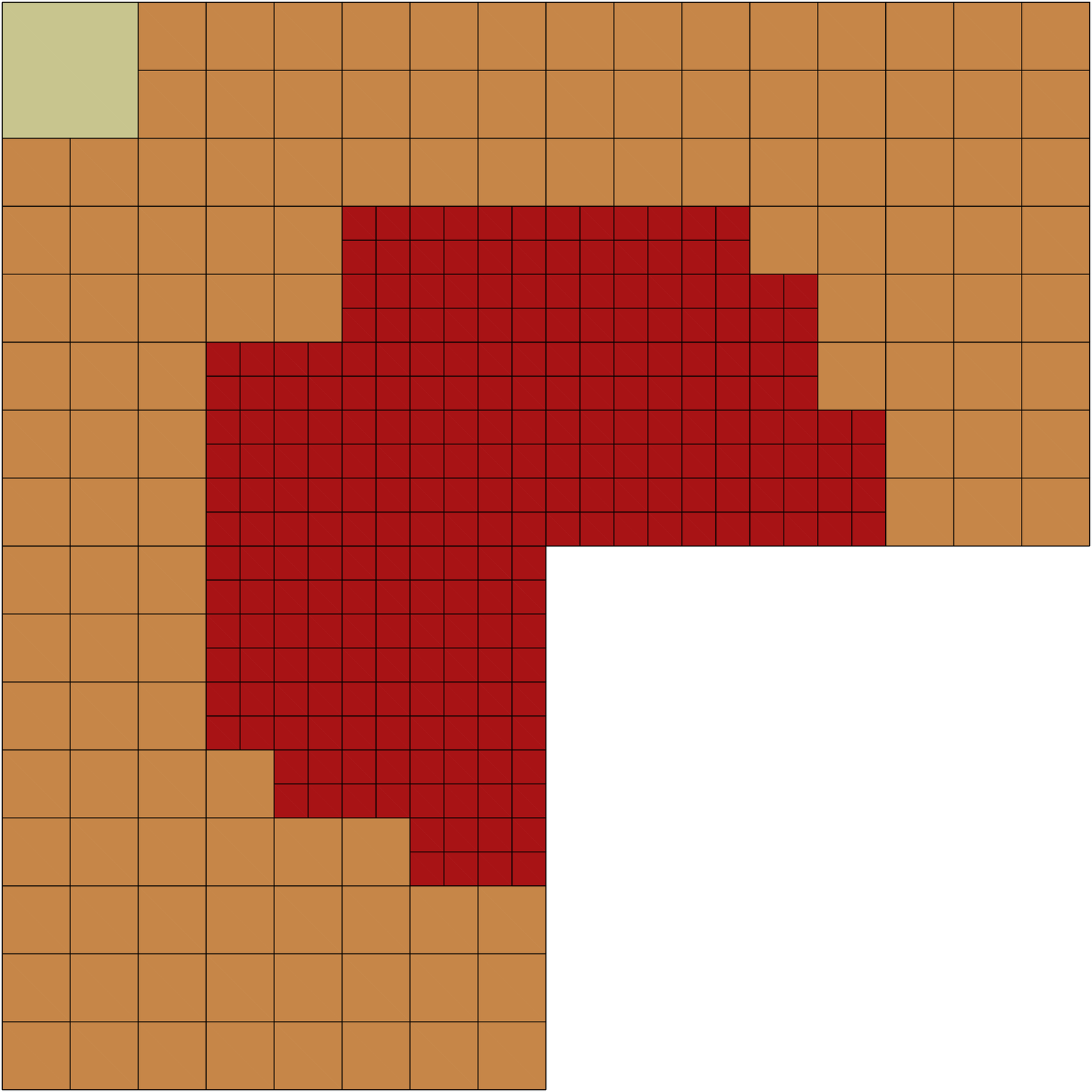}
    \includegraphics[width=0.49\linewidth]{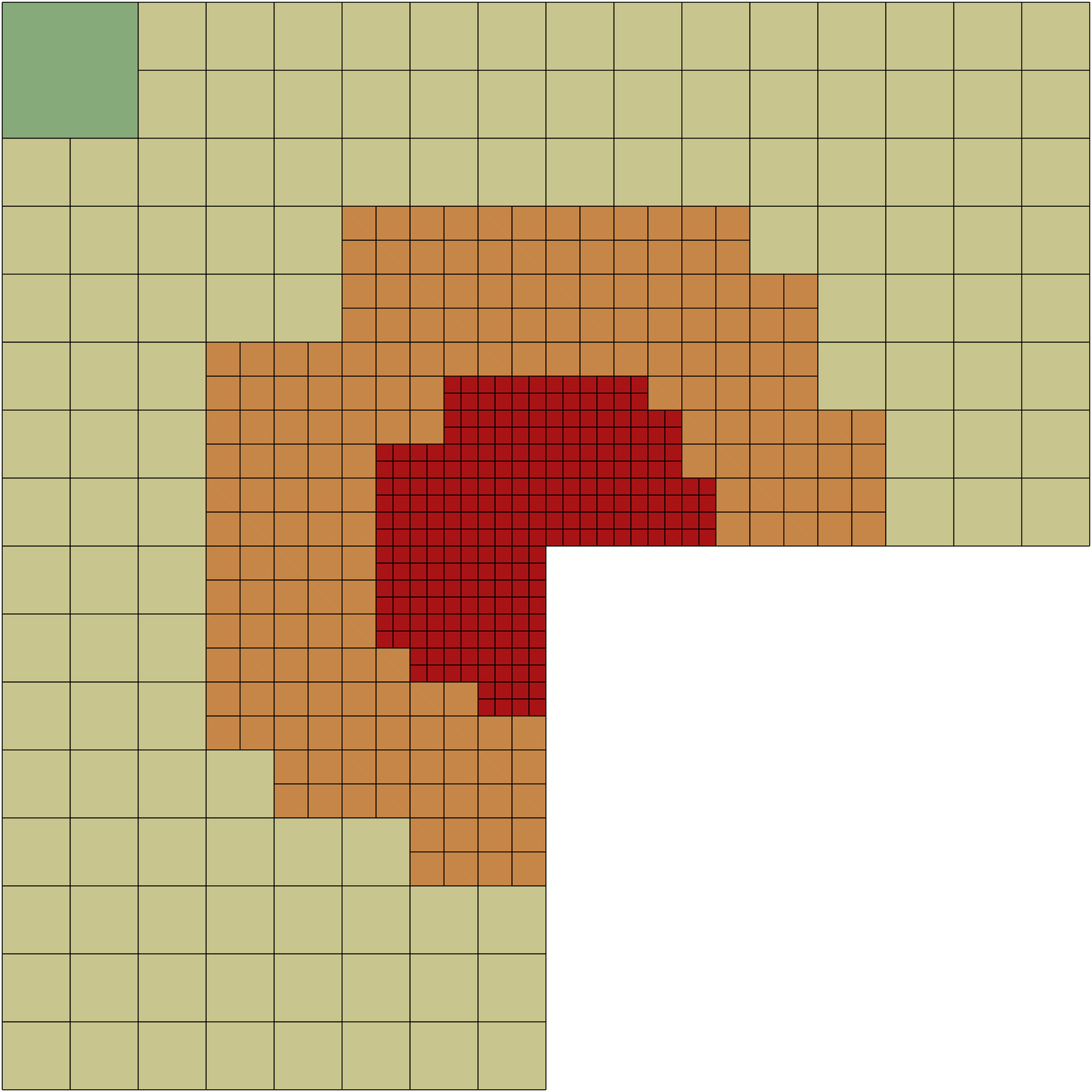} \\
    \includegraphics[width=0.49\linewidth]{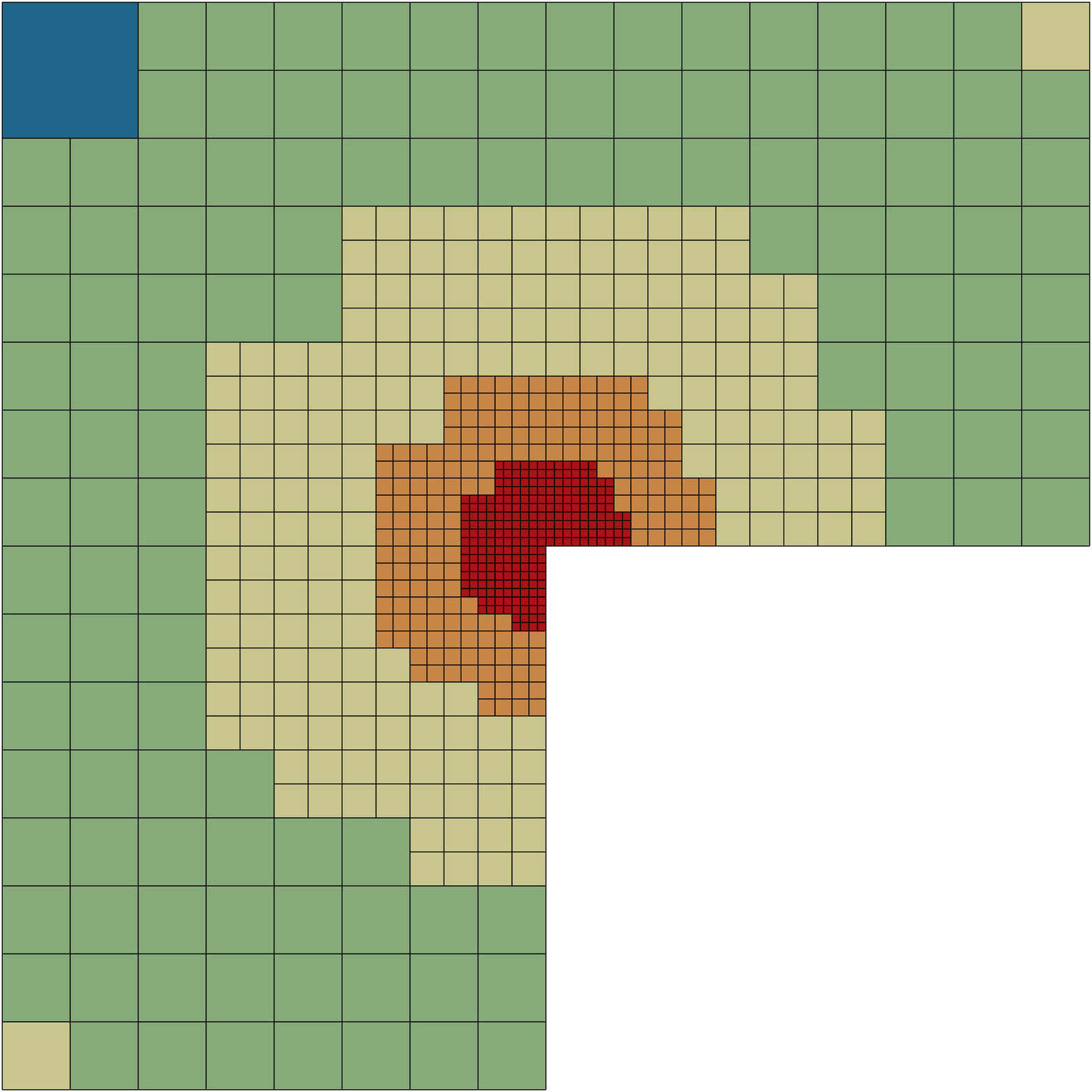}
    \includegraphics[width=0.49\linewidth]{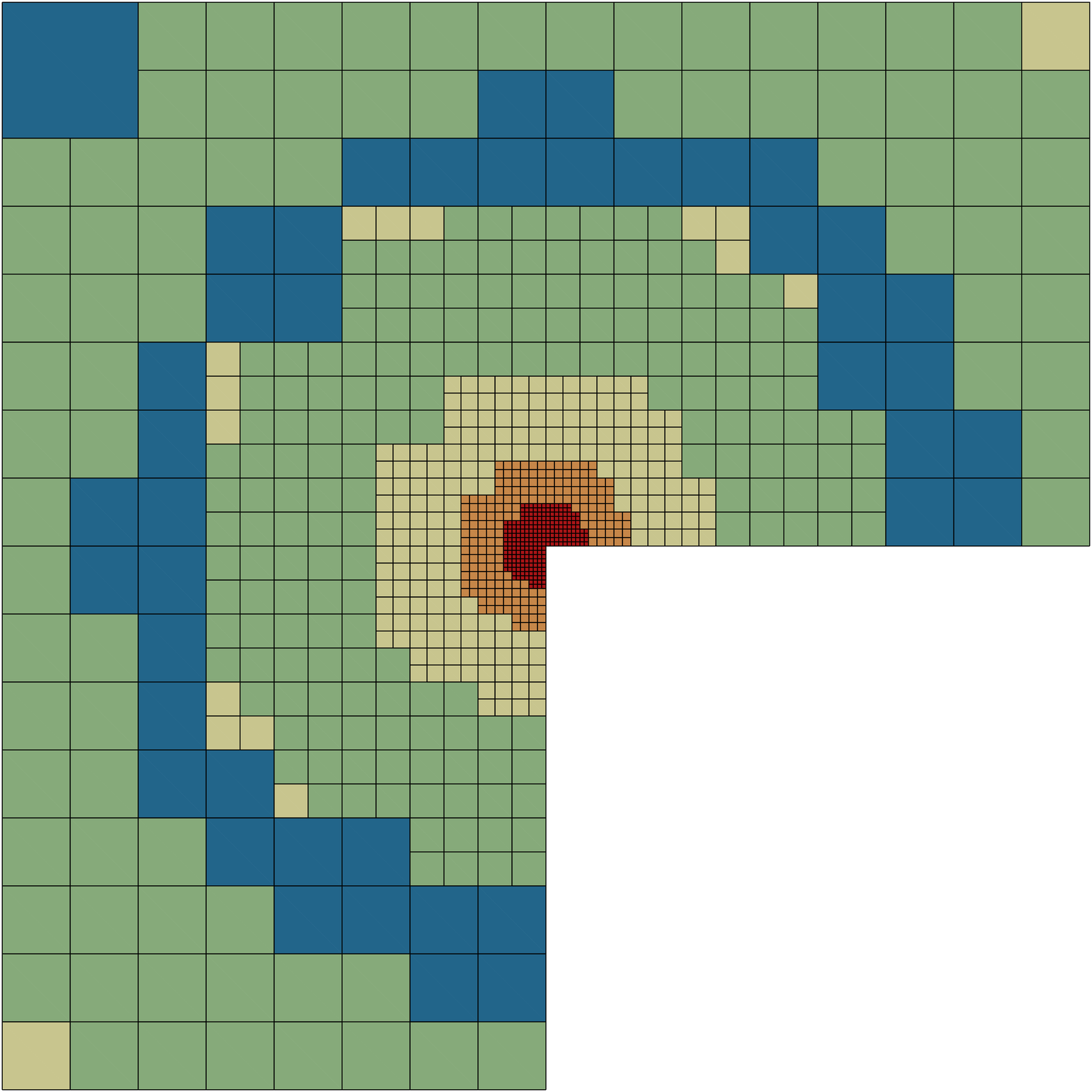} \\
    \includegraphics[width=0.6\linewidth]{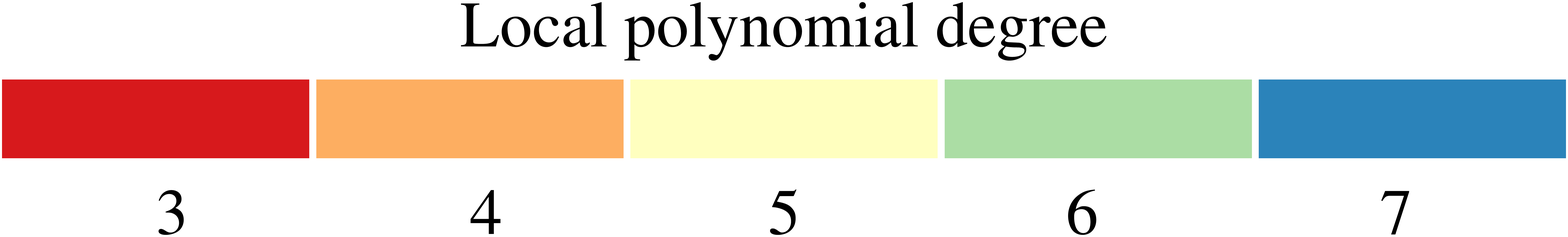}
  \end{minipage}
  \caption{$hp$-meshes generated by the $hp$-adaptive SIPG method when
  solving the reentrant corner benchmark problem on a quadrilateral grid. }
  \label{fig:hp_sipg_quadrilateral_1}
\end{figure}

\begin{figure}[p]
  \centering
  \begin{minipage}{0.8\linewidth}
    \centering
    \includegraphics[width=0.49\linewidth]{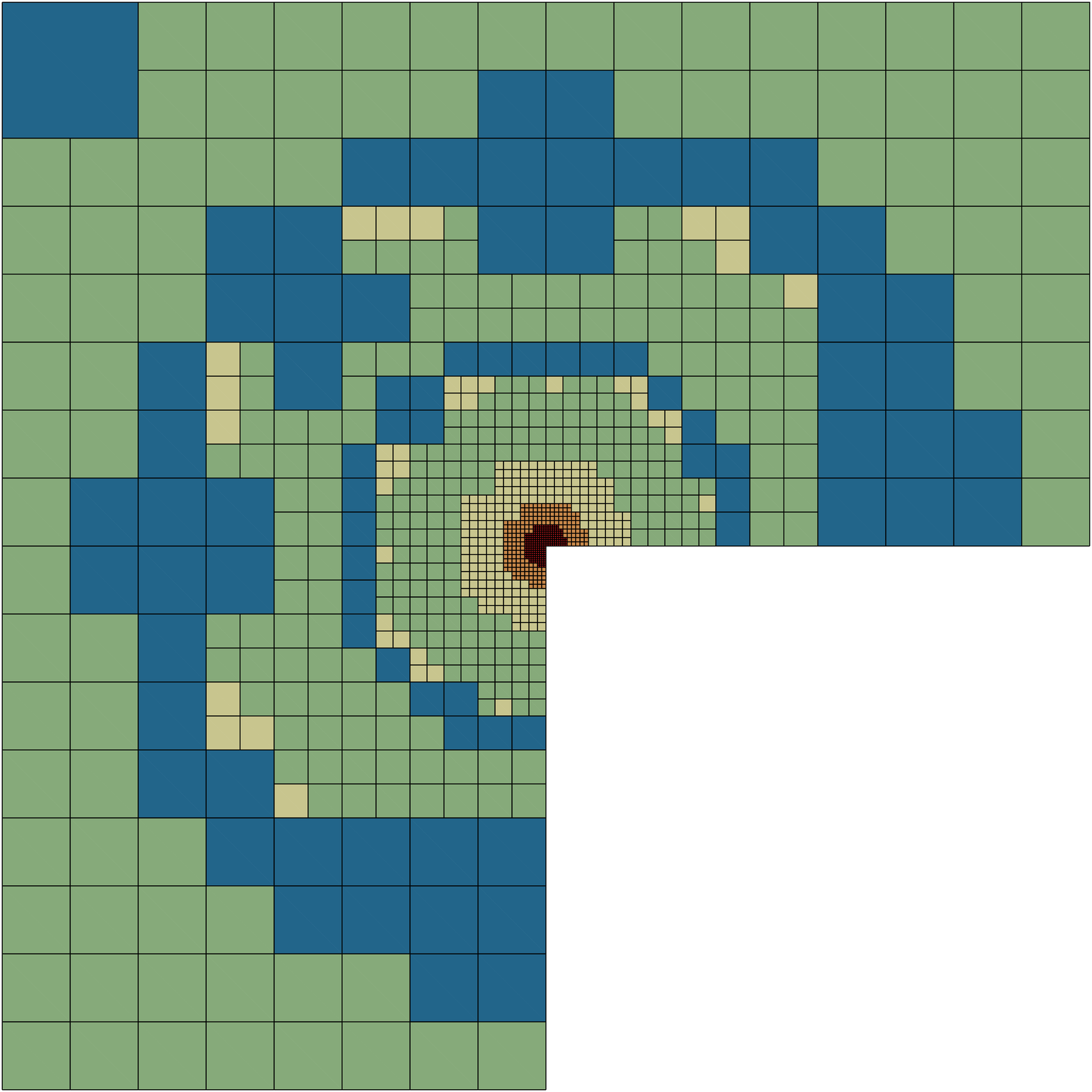}
    \includegraphics[width=0.49\linewidth]{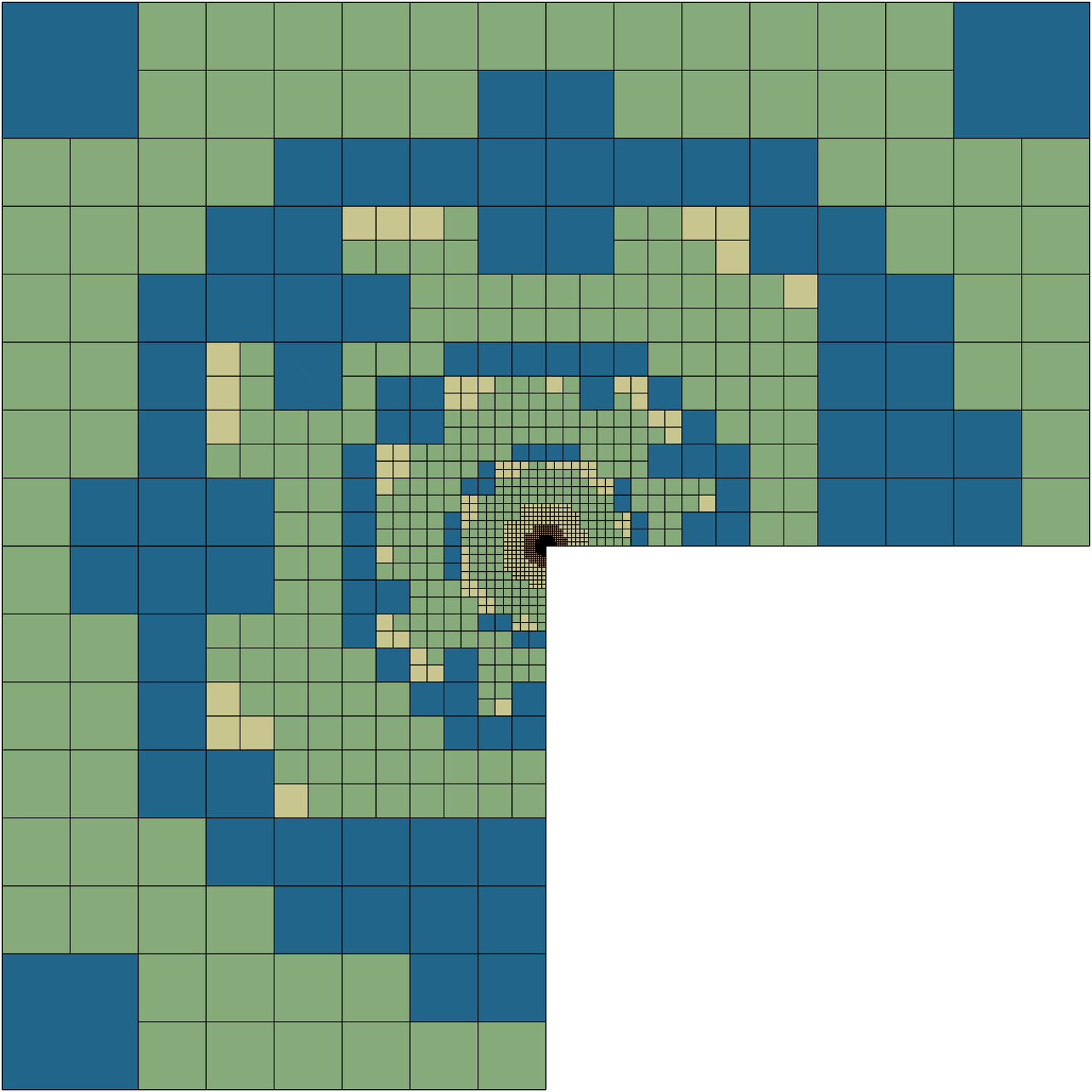} \\
    \includegraphics[width=0.49\linewidth]{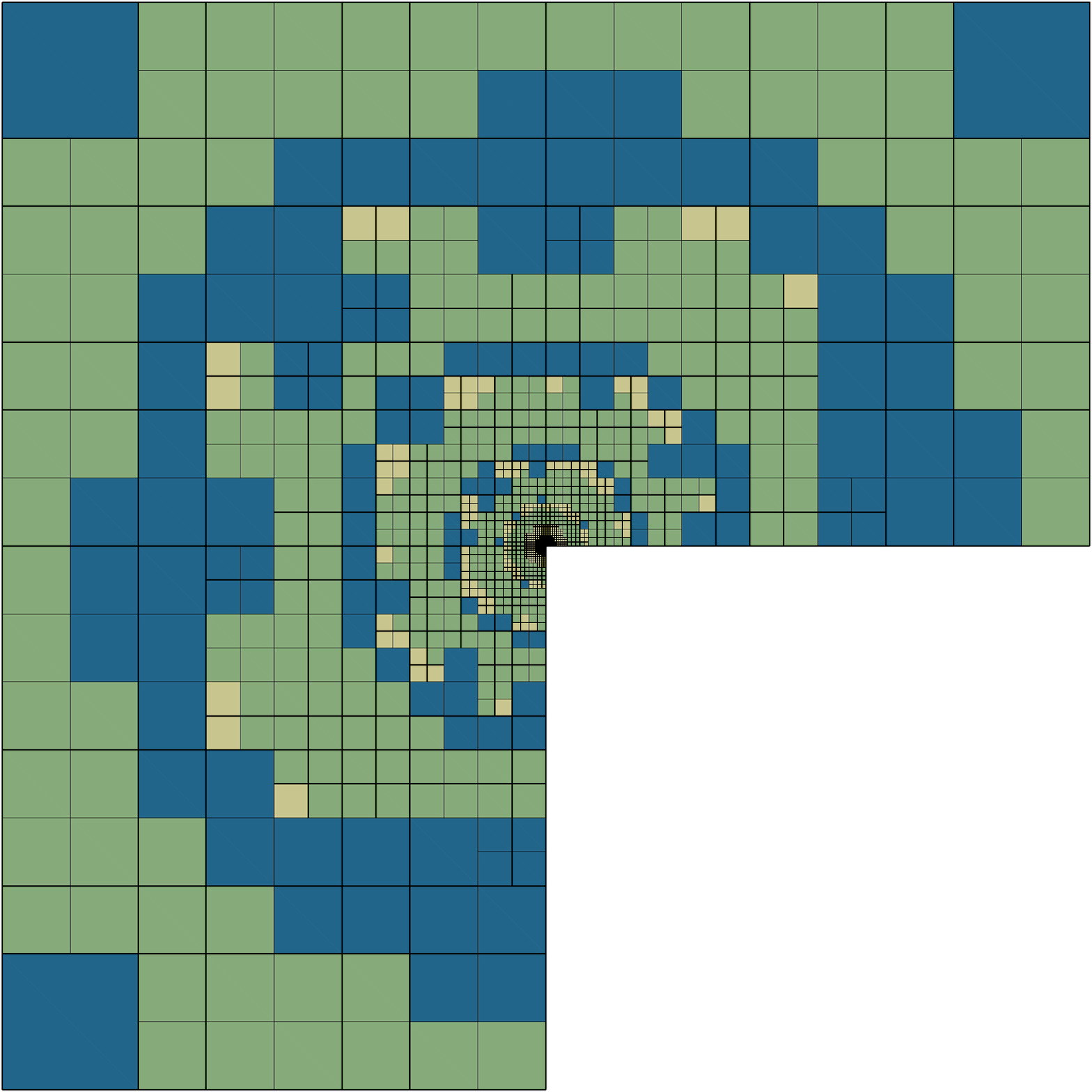} \\
    \includegraphics[width=0.6\linewidth]{colorbar}
  \end{minipage}
  \caption{$hp$-meshes generated by the $hp$-adaptive SIPG method when
  solving the reentrant corner benchmark problem on a quadrilateral grid.}
  \label{fig:hp_sipg_quadrilateral_2}
\end{figure}

We assume that \dune has been configured using CMake. By default, binary
executables will be built out-of-source, e.g., in a designated build
directory \texttt{build\_cmake}. The sources for the $hp$-adaptive SIPG method
can be found in the subdirectory \texttt{examples/poisson}. In order to
verify the results shown in \mbox{Figures
\ref{fig:hp_sipg_quadrilateral_1}, \ref{fig:hp_sipg_quadrilateral_2}} and
\mbox{Table \ref{tbl:hp_sipg_quadrilateral}} compile and run the test as
follows
\begin{lstlisting}[style=display]
cd $DUNE/dune-fem-hpdg/build-cmake
cd ./examples/poisson
make poisson_alugrid_cube_7
./poisson_alugrid_cube_7 ./reentrantcorner.dgf
\end{lstlisting}

\begin{figure}[p]
  \centering
  \begin{minipage}{0.8\linewidth}
    \centering
    \includegraphics[width=0.49\linewidth]{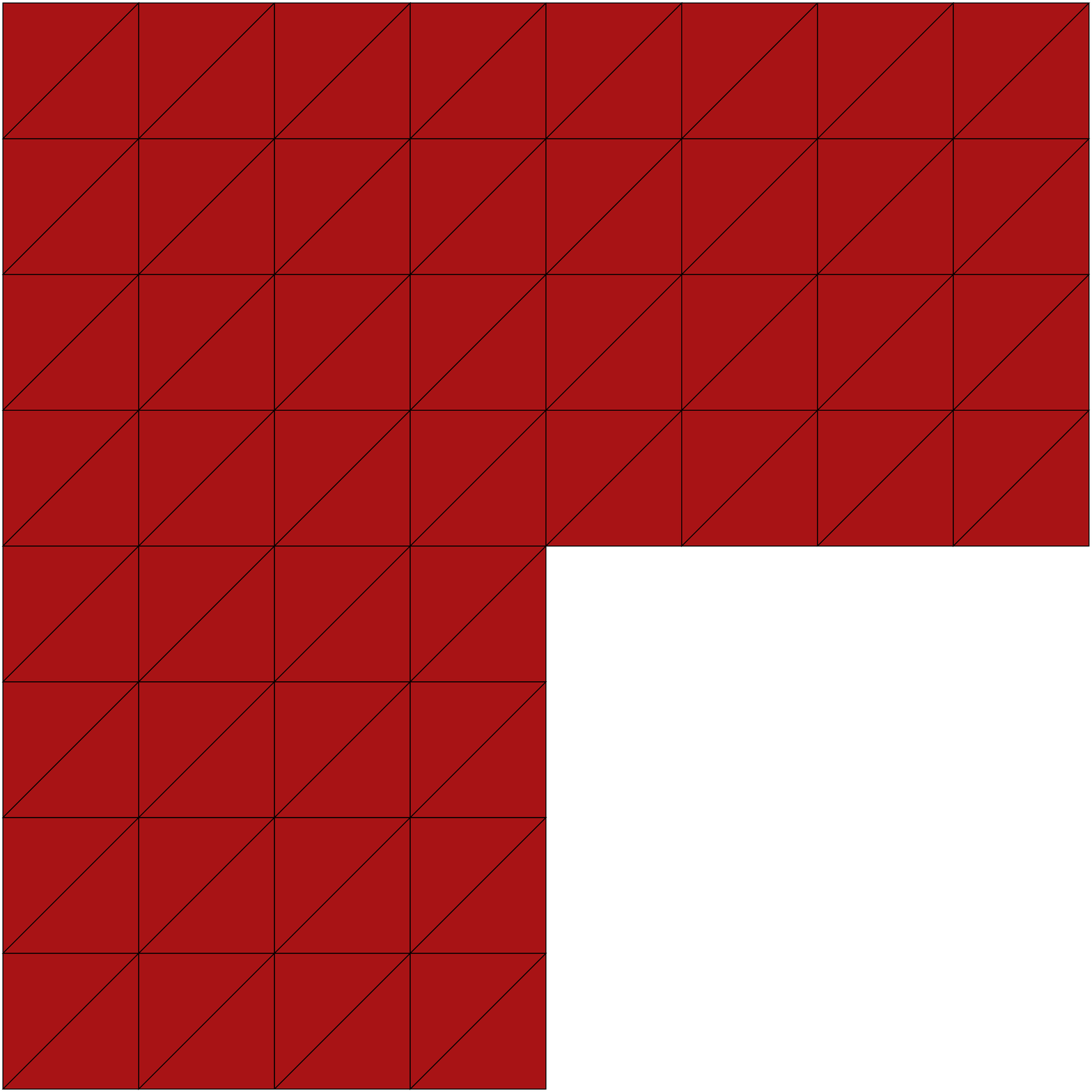}
    \includegraphics[width=0.49\linewidth]{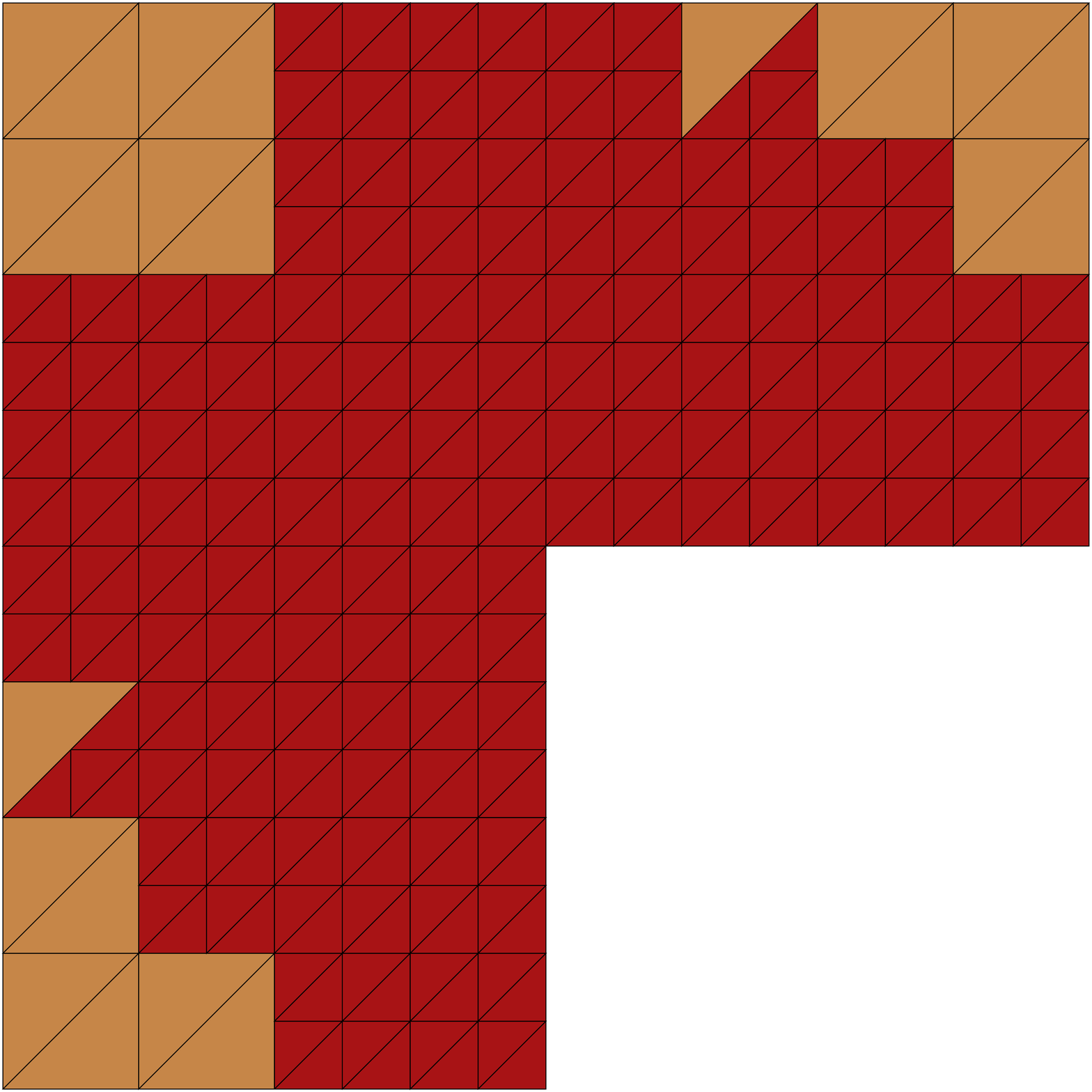} \\
    \includegraphics[width=0.49\linewidth]{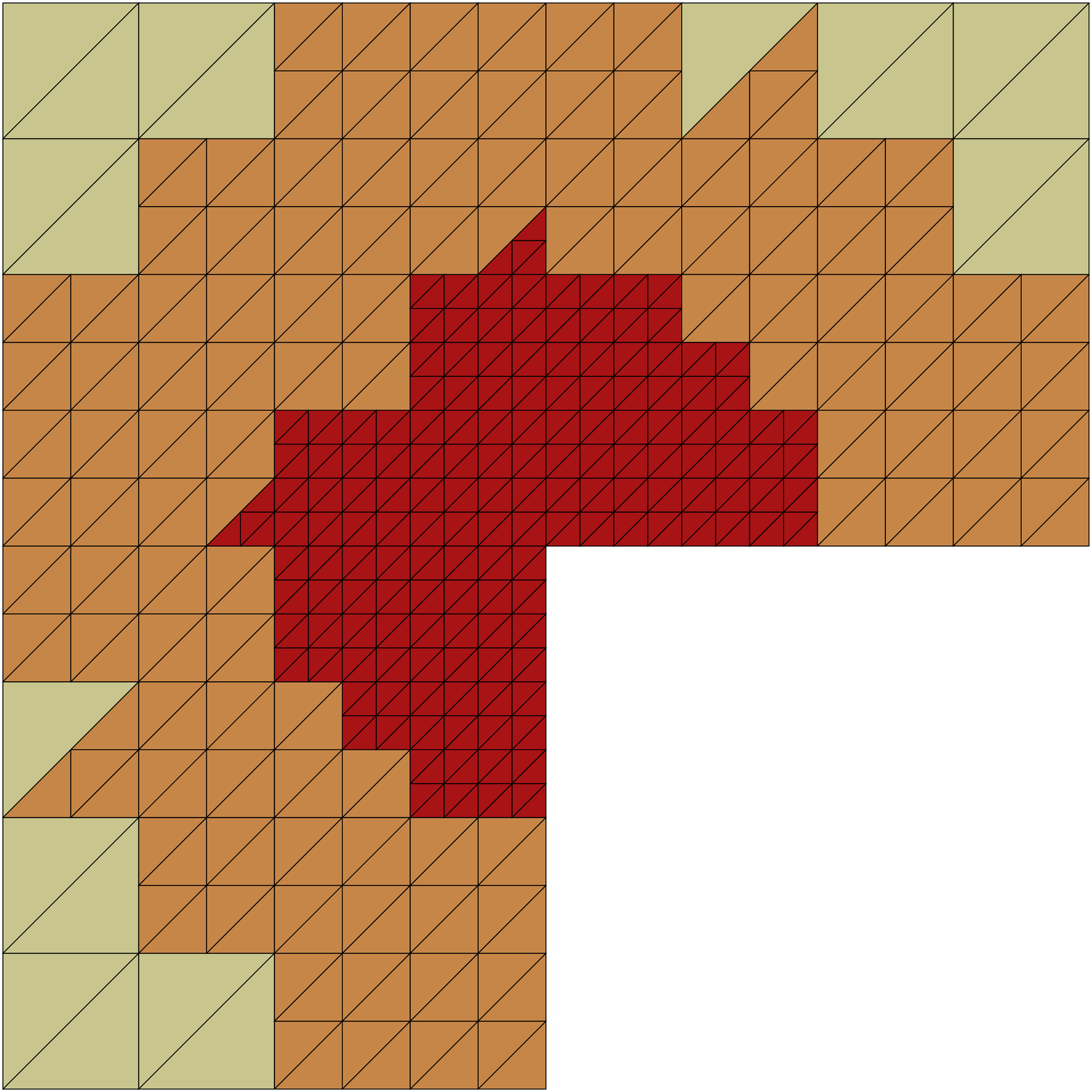}
    \includegraphics[width=0.49\linewidth]{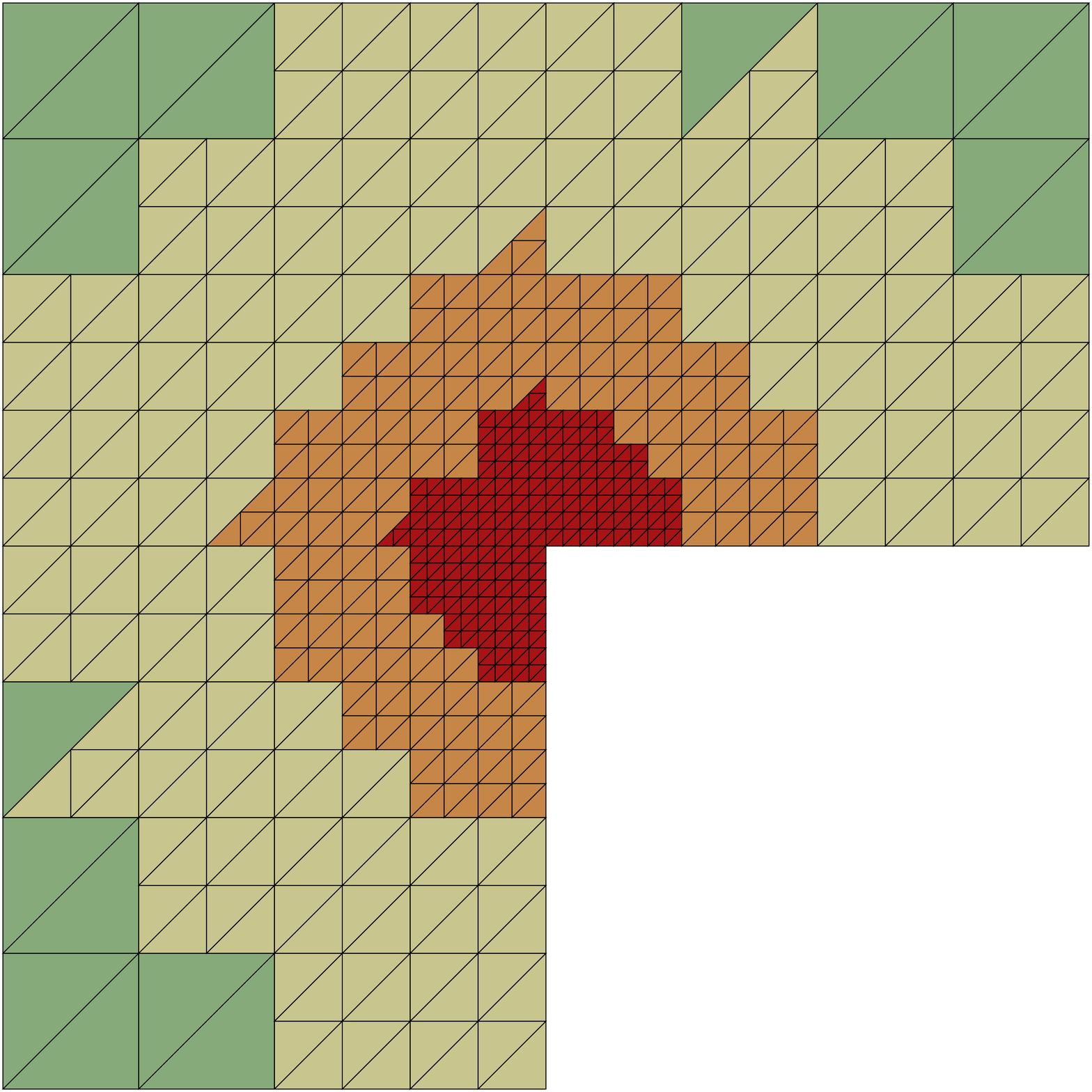} \\
    \includegraphics[width=0.49\linewidth]{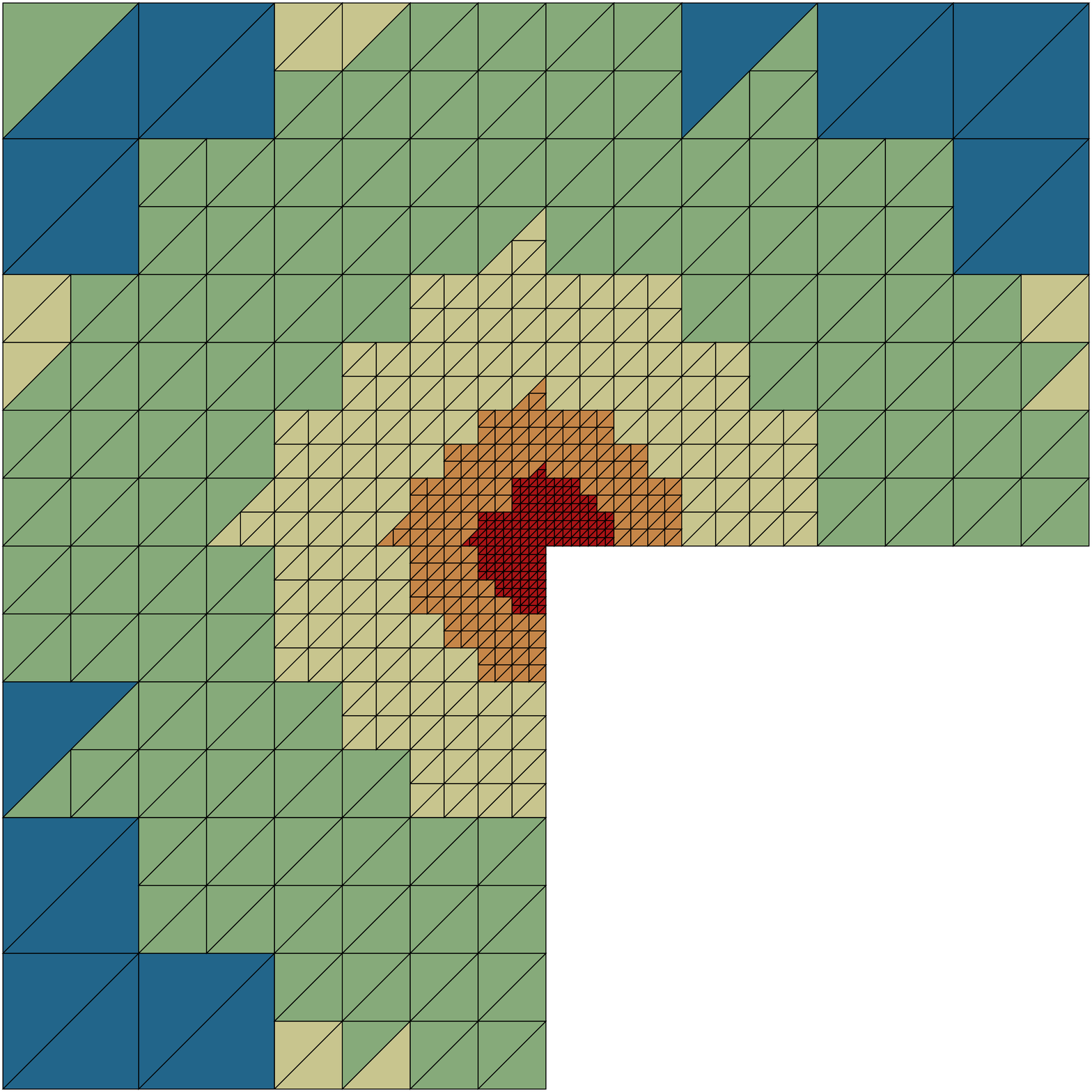}
    \includegraphics[width=0.49\linewidth]{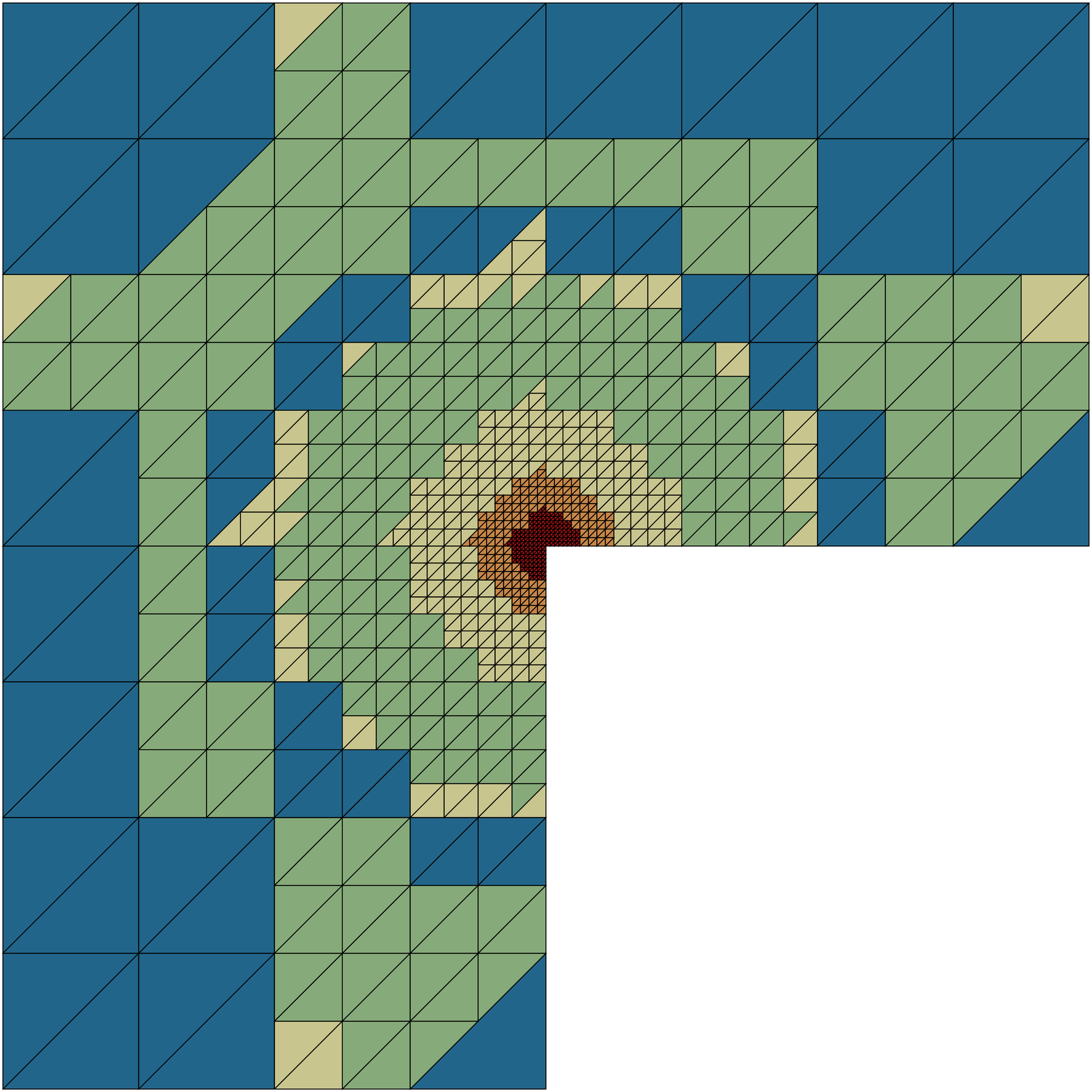} \\
    \includegraphics[width=0.6\linewidth]{colorbar}
  \end{minipage}
  \caption{$hp$-meshes generated by the $hp$-adaptive SIPG method when
  solving the reentrant corner benchmark problem on an triangular grid.}
  \label{fig:hp_sipg_triangular_1}
\end{figure}

\begin{figure}[p]
  \centering
  \begin{minipage}{0.8\linewidth}
    \centering
    \includegraphics[width=0.49\linewidth]{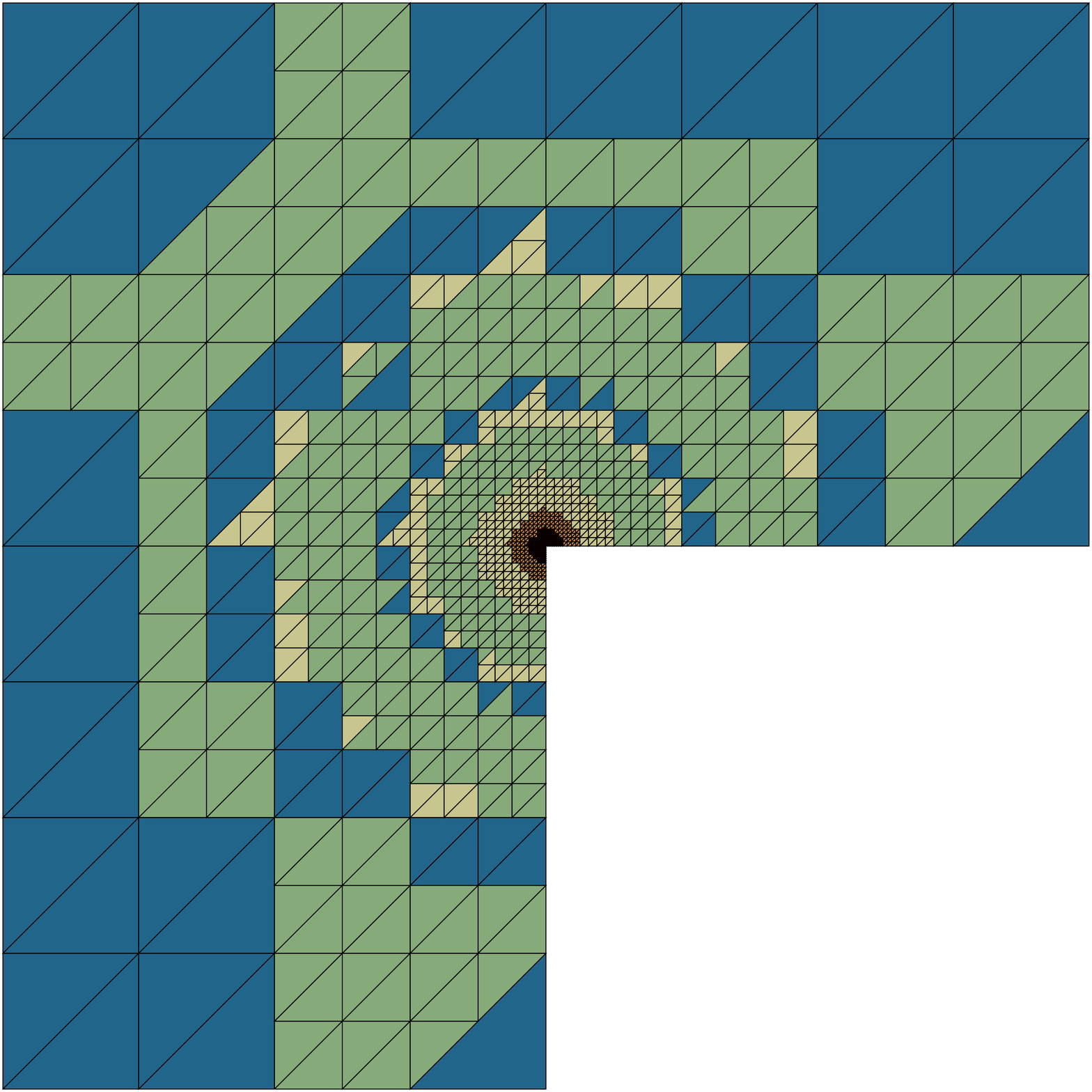}
    \includegraphics[width=0.49\linewidth]{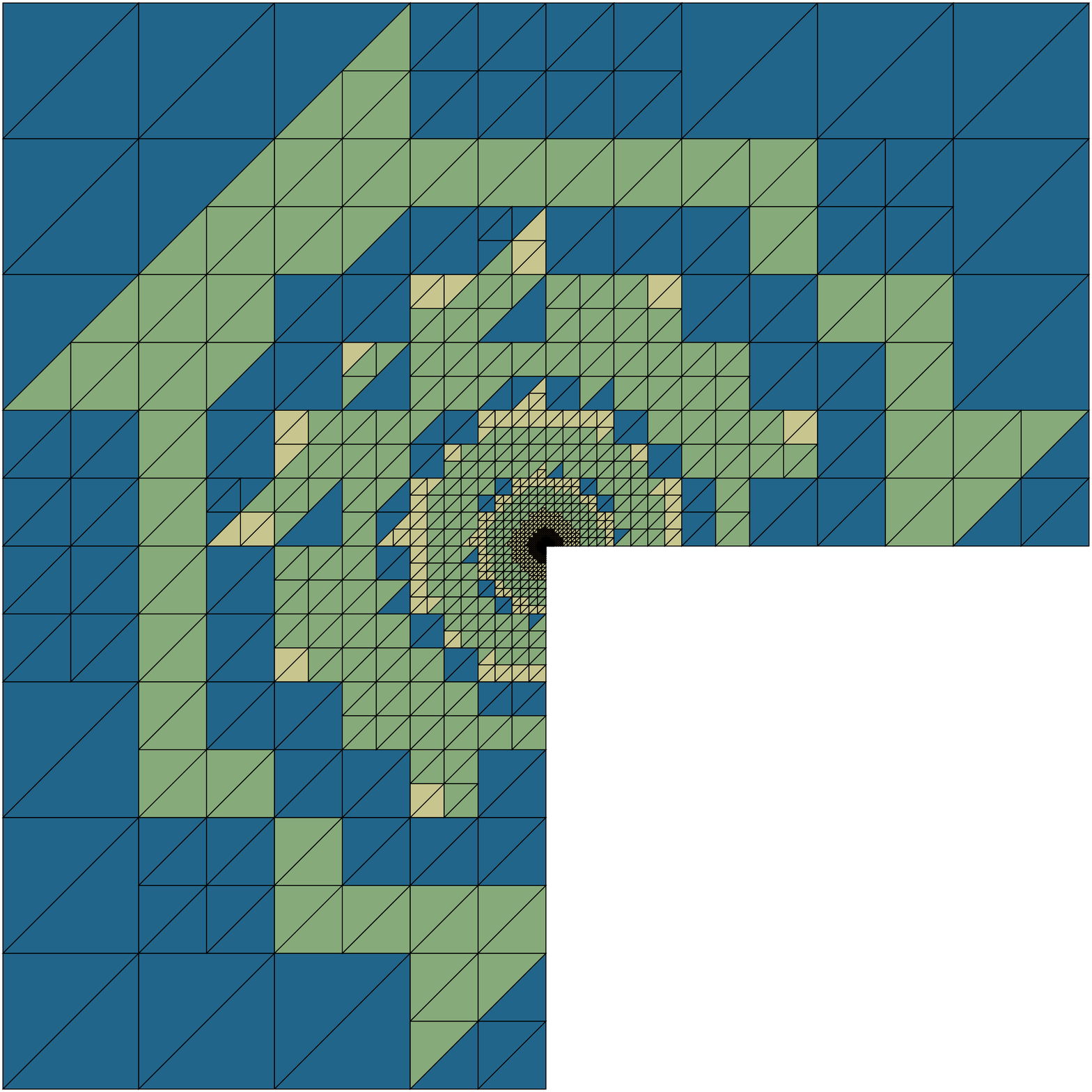} \\
    \includegraphics[width=0.49\linewidth]{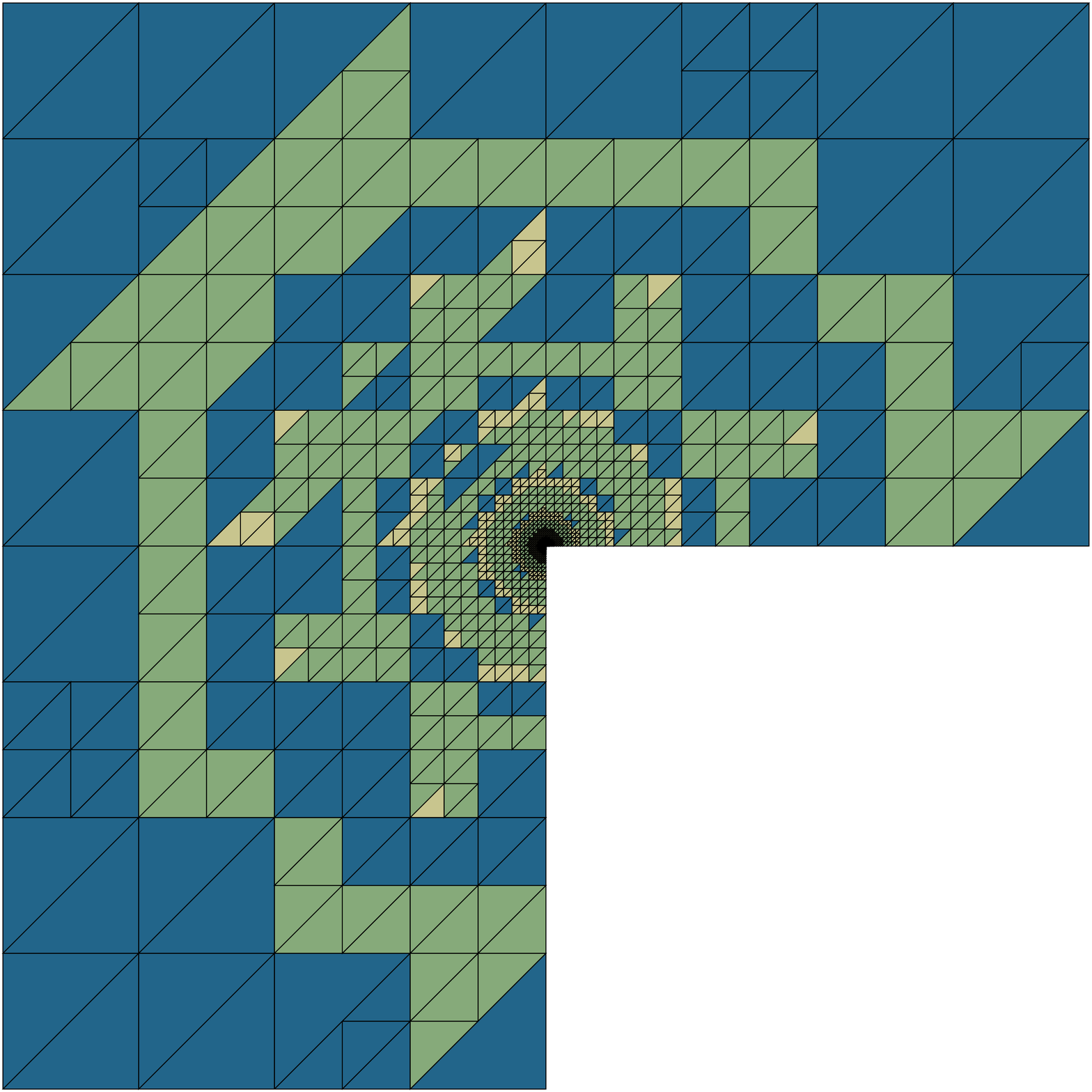} \\
    \includegraphics[width=0.6\linewidth]{colorbar}
  \end{minipage}
  \caption{$hp$-meshes generated by the $hp$-adaptive SIPG method when
  solving the reentrant corner benchmark problem on an triangular grid.}
  \label{fig:hp_sipg_triangular_2}
\end{figure}

The results for the triangular grid shown in \mbox{Figures
\ref{fig:hp_sipg_triangular_1}, \ref{fig:hp_sipg_triangular_2}} and
\mbox{Table \ref{tbl:hp_sipg_triangular}} may be reproduced from running
\begin{lstlisting}[style=display]
make poisson_alugrid_simplex_7
./poisson_alugrid_simplex_7 ./reentrantcorner.dgf
\end{lstlisting}
If an MPI implementation was found during the build process the tests may
be run in parallel in an analogous way.


\section{Advanced user API}
\label{apx:user_api}

In this final appendix we want to describe two advanced aspects of the
\texttt{dune-fem-hpdg} module, the definition of custom data projections for
the restriction and prolongation of user data and the implementation of new
discontinuous finite element spaces based on user-defined local basis
function sets.

\subsection{Restriction and prolongation of custom user data}

Remember from \mbox{Section \ref{sct:dune_fem_hpdg_p_adaption}} that the
restriction and prolongation of data in \texttt{dune-fem-hpdg} is handled by
the \lstinline!hpDG::AdaptationManager! class. Any input must be encapsulated in
a \lstinline!DataProjection! object.
\begin{lstlisting}[style=display]
using DataProjection = hpDG::DefaultDataProjection<Data>;
hpDG::AdaptationManager<DiscreteFunctionSpace, DataProjection> p_AdaptManager(
    space, DataProjection(u));
\end{lstlisting}
A \lstinline!DataProjection! is a callable object encapsulating the data
$u^{(m)}$ associated with the discrete function space $X^{(m)}$. Any such
projection must inherit from the base class \lstinline!hpDG::DataProjection!, as
illustrated by the following listing.
\begin{lstlisting}[style=display]
template <class DiscreteFunctionSpace, class Data>
class DefaultDataProjection
    : public DataProjection<DiscreteFunctionSpace,
                            DefaultDataProjection<Data> > {
 public:
  using BasisFunctionSetType =
      typename DiscreteFunctionSpace::BasisFunctionSetType;
  using EntityType = typename BasisFunctionSetType::EntityType;

  void DataProjection::operator()(const EntityType &element,
                                  const BasisFunctionSetType &former,
                                  const BasisFunctionSetType &future,
                                  const vector<size_t> &origin,
                                  const vector<size_t> &destination);
};
\end{lstlisting}
The method \lstinline!operator()! is expected to evaluate the local projection
$\Pi^{(m+1)}u^{(m)}\restr{E}$ for each element $E \in \grid^{(m)} \cap
\grid^{(m+)1}$. Its input arguments are
\vspace{-\topsep}
\begin{enumerate}
  \item the grid element $E$,

  \item the former and future local basis function sets
  \begin{math}
    \left(\phi_{E,i}^{(m)}\right)_{i = 0, \ldots, n_E^{(m)}-1},~
    \left(\phi_{E,i}^{(m+1)}\right)_{i = 0, \ldots, n_E^{(m+1)}-1},
  \end{math}

  \item and the DOF indices
  \begin{math}
    (\mu^{(m)}(E, i))_{i = 0, \ldots, n_E^{(m)}-1},~
    (\mu^{(m+1/2)}(E, i))_{i = 0, \ldots, n_E^{(m+1)}-1}.
  \end{math}
\end{enumerate}
\vspace{-\topsep}
Note that this is all information needed in order to implement the local
$L^2$-projection from \mbox{Example \ref{exp:local_projection}}.

\subsection{Setup of new adaptive discontinuous finite element spaces}

\begin{figure}
  \centering
  \includegraphics[width=\linewidth]{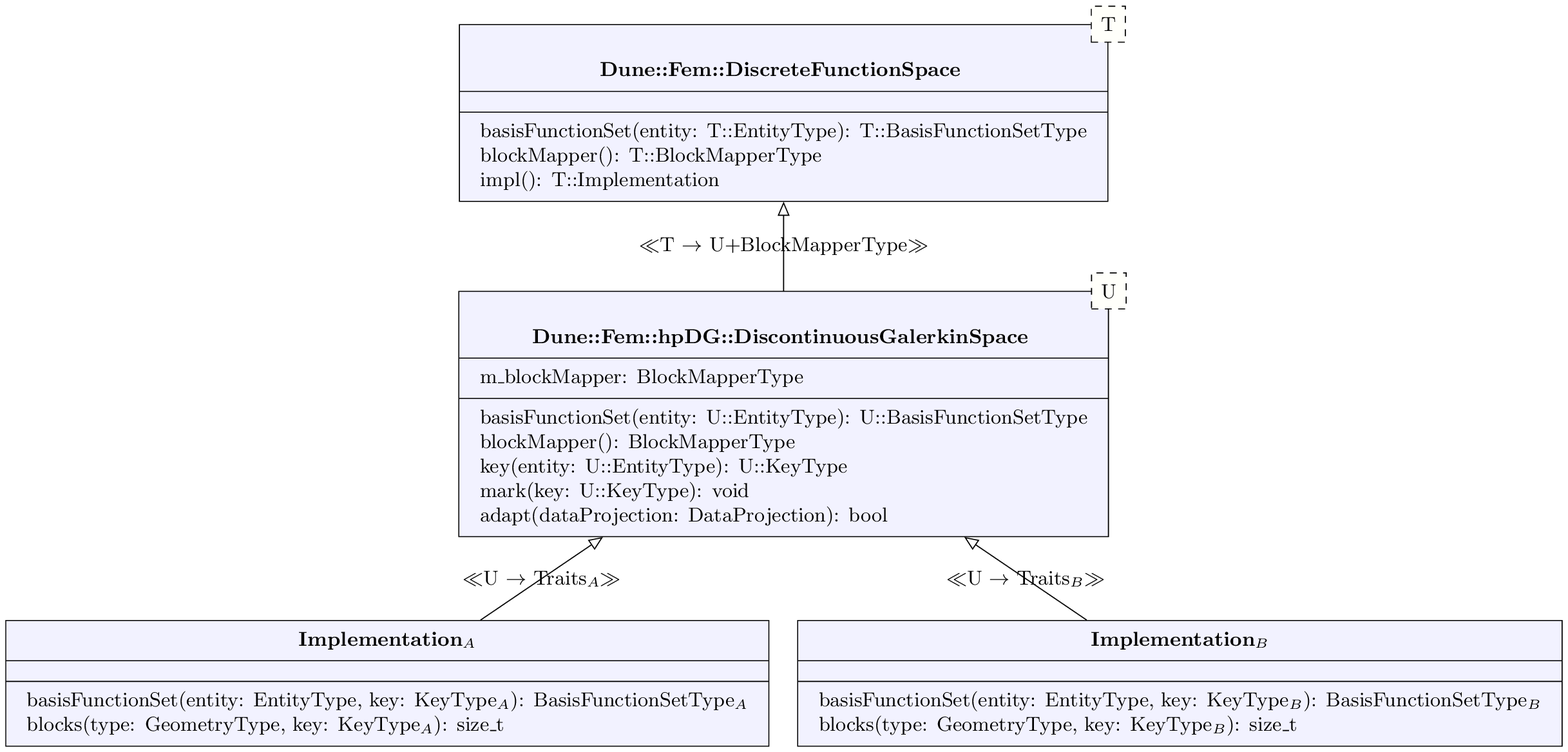}
  \caption{Class diagram for derived implementations of $hp$-adaptive
  discontinuous finite element spaces inheriting from the
  \lstinline!hpDG::DiscontinuousGalerkinSpace! base class.}
  \label{fig:uml_diagram}
\end{figure}

As described in \mbox{Section \ref{sct:dune_fem_hpdg_spaces}} the discrete
function spaces provided by \texttt{dune-fem-hpdg} inherit from a common base
class implementing almost all functionality, see \mbox{Figure
\ref{fig:uml_diagram}}. In the same way, users may setup new discontinuous
finite element spaces. To do so, users need to provide an implementation of
the \lstinline!hpDG::BasisFunctionSets! interface which contains essentially two
methods.

\begin{lstlisting}[style=display]
template <class BasisFunctionSet, class Key>
class BasisFunctionSets {
 public:
  using EntityType = typename BasisFunctionSet::EntityType;

  BasisFunctionSet basisFunctionSet(const EntityType &entity,
                                    const Key &key) const;
  size_t blocks(GeometryType type, const Key &key) const;
  // ...
};
\end{lstlisting}
Remember that a \lstinline!GeometryType! is a \dune data type that identifies a
reference element. Given a grid element $E$ with reference element $R_E$
and a key $k$ the method \lstinline!basisFunctionSet! returns the corresponding
local basis function set $\basis_{E, k}$. We assume that the size of the
local basis function set only depends on $R_E$ and $k$. Then, the method
\lstinline!blocks! returns the number of blocks to be reserved for $\basis_{E,
k}$. A \lstinline!BlockMapper! is a \dune[Fem] class representation a DOF
mapping; in case of a scalar function spaces the number of blocks coincides
with the number of local basis functions in $\basis_{E, k}$.

Users may inherit from the base implementation
\lstinline!hpDG::DiscontinuousGalerkinSpace!. Alternatively, the convenience
function \lstinline!make_space! immediately constructs a discrete function space,
given a grid part, the family of basis function sets, and a default key for
initializing the space.
\begin{lstlisting}[style=display]
template <class GridPart, class BasisFunctionSets>
std::unique_ptr<DefaultDiscontinuousGalerkinSpace<BasisFunctionSets> >
make_space(GridPart &gridPart, const BasisFunctionSets &basisFunctionSets,
           const typename BasisFunctionSets::KeyType &key);
\end{lstlisting}



\begin{table}
  \centering
  \caption{Number of elements and degrees of freedom, errors and
  convergence rates, and effectivity index for the series of quadrilateral
  grids depicted in \mbox{Figures \ref{fig:hp_sipg_quadrilateral_1}} and
  \ref{fig:hp_sipg_quadrilateral_2}.}
  \label{tbl:hp_sipg_quadrilateral}
  \begin{tabular}{
     S[table-format=4.0]
     S[table-format=5.0]
     S[table-format=1.2e-1]
     S[table-format=2.1]
     S[table-format=1.2e-1]
     S[table-format=1.1]
     S[table-format=1.2e-1]
     S[table-format=1.2]
   }
  \toprule
  {Elements} & {DoFs} & {$\norm{u - u_h}_{L^2}$} & {EOC} & {$\norm{u - u_h}_{\mathit{DG}}$} & {EOC} & {$\left(\sum \eta_E^2\right)^{\frac{1}{2}}$} & {Eff. index} \\
  \midrule
  48 & 480 & 2.30e-03 & {---} & 2.22e-01 & {---} & 3.30e-01 & 1.49 \\
  189 & 1895 & 8.56e-04 & 1.4 & 1.39e-01 & 0.7 & 2.08e-01 & 1.49 \\
  378 & 4416 & 3.25e-04 & 2.3 & 8.77e-02 & 1.1 & 1.31e-01 & 1.49 \\
  567 & 8008 & 1.25e-04 & 3.2 & 5.52e-02 & 1.6 & 8.23e-02 & 1.49 \\
  756 & 12846 & 4.85e-05 & 4.0 & 3.48e-02 & 2.0 & 5.19e-02 & 1.49 \\
  945 & 18376 & 1.90e-05 & 5.2 & 2.19e-02 & 2.6 & 3.27e-02 & 1.49 \\
  1104 & 22964 & 7.47e-06 & 8.4 & 1.38e-02 & 4.1 & 2.06e-02 & 1.49 \\
  1269 & 27480 & 2.95e-06 & 10.4 & 8.70e-03 & 5.1 & 1.30e-02 & 1.49 \\
  1464 & 32712 & 1.17e-06 & 10.6 & 5.48e-03 & 5.3 & 8.17e-03 & 1.49 \\
  \bottomrule
  \end{tabular}
\end{table}

\begin{table}
  \centering
  \caption{Number of elements and degrees of freedom, errors and
  convergence rates, and effectivity index for the series of triangular
  grids depicted in \mbox{Figures \ref{fig:hp_sipg_triangular_1}} and
  \ref{fig:hp_sipg_triangular_2}.}
  \label{tbl:hp_sipg_triangular}
  \begin{tabular}{
     S[table-format=4.0]
     S[table-format=5.0]
     S[table-format=1.2e-1]
     S[table-format=2.1]
     S[table-format=1.2e-1]
     S[table-format=1.1]
     S[table-format=1.2e-1]
     S[table-format=1.2]
   }
  \toprule
  {Elements} & {DoFs} & {$\norm{u - u_h}_{L^2}$} & {EOC} & {$\norm{u - u_h}_{\mathit{DG}}$} & {EOC} & {$\left(\sum \eta_E^2\right)^{\frac{1}{2}}$} & {Eff. index} \\
  \midrule
  96 & 960 & 1.08e-03 & {---} & 2.31e-01 & {---} & 3.04e-01 & 1.32 \\
  318 & 3290 & 4.09e-04 & 1.6 & 1.44e-01 & 0.8 & 1.92e-01 & 1.33 \\
  558 & 6930 & 1.57e-04 & 2.6 & 9.06e-02 & 1.2 & 1.21e-01 & 1.33 \\
  798 & 12014 & 6.08e-05 & 3.4 & 5.70e-02 & 1.7 & 7.60e-02 & 1.33 \\
  1038 & 18690 & 2.37e-05 & 4.3 & 3.59e-02 & 2.1 & 4.79e-02 & 1.33 \\
  1227 & 24137 & 9.33e-06 & 7.3 & 2.26e-02 & 3.6 & 3.02e-02 & 1.34 \\
  1464 & 30643 & 3.68e-06 & 7.8 & 1.42e-02 & 3.9 & 1.90e-02 & 1.34 \\
  1716 & 37871 & 1.46e-06 & 8.8 & 8.97e-03 & 4.4 & 1.20e-02 & 1.34 \\
  1899 & 42361 & 5.76e-07 & 16.5 & 5.65e-03 & 8.2 & 7.54e-03 & 1.34 \\
  \bottomrule
  \end{tabular}
\end{table}


\begin{thebibliography}{10}

\bibitem{Alkaemper2016}
M.~Alk{\"a}mper, A.~Dedner, R.~Kl{\"o}fkorn, and M.~Nolte.
\newblock The \textsc{Dune-{ALU}Grid }module.
\newblock {\em {Arch. Numer. Softw.}}, 4(1):1--28, 2016.

\bibitem{Babuska1981}
I.~Babu{\v{s}}ka, B.~A. Szabo, and I.~N. Katz.
\newblock The {$p$}-version of the finite element method.
\newblock {\em {SIAM J. Numer. Anal.}}, 18(3):515--545, 1981.

\bibitem{Bangerth2007}
W.~Bangerth, R.~Hartmann, and G.~Kanschat.
\newblock {deal.II} -- {A} general purpose object oriented finite element
  library.
\newblock {\em {ACM Trans. Math. Softw.}}, 33(4):24/1--24/27, 2007.

\bibitem{Bangerth2009}
W.~Bangerth and O.~Kayser-Herold.
\newblock Data structures and requirements for {$hp$} finite element software.
\newblock {\em {ACM Trans. Math. Softw.}}, 36(1):4:1--4:31, 2009.

\bibitem{Bastian2008a}
P.~Bastian, M.~Blatt, A.~Dedner, {\relax Ch}.~Engwer, R.~Kl{\"o}fkorn,
  R.~Kornhuber, M.~Ohlberger, and O.~Sander.
\newblock A generic grid interface for parallel and adaptive scientific
  computing. {Part II}: {Implementation} and tests in \textsc{Dune}.
\newblock {\em {Computing}}, 82(2--3):121--138, 2008.

\bibitem{Bastian2008}
P.~Bastian, M.~Blatt, A.~Dedner, {\relax Ch}.~Engwer, R.~Kl{\"o}fkorn,
  M.~Ohlberger, and O.~Sander.
\newblock A generic grid interface for parallel and adaptive scientific
  computing. {Part I}: {Abstract} framework.
\newblock {\em {Computing}}, 82(2--3):103--119, 2008.

\bibitem{Blatt2007}
M.~Blatt and P.~Bastian.
\newblock The iterative solver template library.
\newblock In B.~K\aa{}gstr{\"o}m, E.~Elmroth, J.~Dongarra, and
  J.~Wa{\'s}niewski, editors, {\em Applied Parallel Computing. {State} of the
  Art in Scientific Computing}, pages 666--675. Springer Berlin Heidelberg,
  2007.

\bibitem{Blatt2008}
M.~Blatt and P.~Bastian.
\newblock On the generic parallelisation of iterative solvers for the finite
  element method.
\newblock {\em {Int. J. Comput. Sci. Eng.}}, 4(1):56--69, 2008.

\bibitem{Dedner2014}
A.~Dedner, R.~Kl{\"o}fkorn, and M.~Kr{\"a}nkel.
\newblock Continuous finite-elements on non-conforming grids using
  discontinuous {Galerkin} stabilization.
\newblock In J.~Fuhrmann, M.~Ohlberger, and {\relax Ch}.~Rohde, editors, {\em
  Finite Volumes for Complex Applications {VII}. {Methods} and Theoretical
  Aspects}, pages 207--215. Springer International Publishing, 2014.

\bibitem{Dedner2010}
A.~Dedner, R.~Kl{\"o}fkorn, M.~Nolte, and M.~Ohlberger.
\newblock A generic interface for parallel and adaptive discretization schemes:
  {Abstraction} principles and the \textsc{Dune-Fem} module.
\newblock {\em {Computing}}, 90(3--4):165--196, 2010.

\bibitem{Demkowicz2002}
L.~Demkowicz, W.~Rachowicz, and {\relax Ph}.~Devloo.
\newblock A fully automatic {$hp$}-adaptivity.
\newblock {\em {J. Sci. Comput.}}, 17(1):117--142, 2002.

\bibitem{Dolejsi2013}
V.~Dolej\v{s}\'{\i}.
\newblock {$hp$-DGFEM} for nonlinear convection-diffusion problems.
\newblock {\em {Math. Comput. Simulation}}, 87:87--118, 2013.

\bibitem{Dolejsi2015}
V.~Dolej\v{s}\'{\i} and M.~Feistauer.
\newblock {\em Discontinuous {Galerkin} Method. {Analysis} and Applications to
  Compressible Flow.}
\newblock Springer International Publishing, 2015.

\bibitem{Eibner2007}
T.~Eibner and J.~Melenk.
\newblock An adaptive strategy for {$hp$-FEM} based on testing for analyticity.
\newblock {\em {Comput. Mech.}}, 39(5):575--595, 2007.

\bibitem{Frauenfelder2002}
{\relax Ph}.~Frauenfelder and {\relax Ch}.~Lage.
\newblock Concepts - {An} object-oriented software package for partial
  differential equations.
\newblock {\em {ESAIM Math. Model. Numer. Anal.}}, 36(5):937--951, 2002.

\bibitem{Gui1986}
W.~Gui and I.~Babu{\v{s}}ka.
\newblock The {$h$}, {$p$} and {$h$-$p$} versions of the finite element method
  in one dimension. {Part I}. {The} error analysis of the {$p$}-version.
\newblock {\em {Numer. Math.}}, 49(6):577--612, 1986.

\bibitem{Gui1986a}
W.~Gui and I.~Babu{\v{s}}ka.
\newblock The {$h$}, {$p$} and {$h$-$p$} versions of the finite element method
  in one dimension. {Part II}. {The} error analysis of the {$h$}- and {$h$-$p$}
  versions.
\newblock {\em {Numer. Math.}}, 49(6):613--657, 1986.

\bibitem{Gui1986b}
W.~Gui and I.~Babu{\v{s}}ka.
\newblock The {$h$}, {$p$} and {$h$-$p$} versions of the finite element method
  in one dimension. {Part III}. {The} adaptive {$h$-$p$}-version.
\newblock {\em {Numer. Math.}}, 49(6):659--683, 1986.

\bibitem{Guo1986}
B.~Guo and I.~Babu{\v{s}}ka.
\newblock The {$h$-$p$} version of the finite element method. {Part I}. {The}
  basic approximation results.
\newblock {\em {Comput. Mech.}}, 1(1):21--41, 1986.

\bibitem{Guo1986a}
B.~Guo and I.~Babu{\v{s}}ka.
\newblock The {$h$-$p$} version of the finite element method. {Part II}.
  {General} results and applications.
\newblock {\em {Comput. Mech.}}, 1(3):203--220, 1986.

\bibitem{Houston2007}
P.~Houston, D.~Sch{\"o}tzau, and {\relax Th}.~P. Wihler.
\newblock Energy norm a posteriori error estimation of {$hp$}-adaptive
  discontinuous {Galerkin} methods for elliptic problems.
\newblock {\em {Math. Models Methods Appl. Sci.}}, 17(01):33--62, 2007.

\bibitem{Houston2003}
P.~Houston, B.~Senior, and E.~S{\"u}li.
\newblock {\em Numerical Mathematics and Advanced Applications}, chapter
  Sobolev regularity estimation for {$hp$}-adaptive finite element methods,
  pages 631--656.
\newblock Springer Milan, 2003.

\bibitem{Houston2005}
P.~Houston and E.~S{\"u}li.
\newblock A note on the design of {$hp$}-adaptive finite element methods for
  elliptic partial differential equations.
\newblock {\em {Comput. Method. Appl. M.}}, 194(2--5):229--243, 2005.

\bibitem{Houston2008}
P.~Houston, E.~S{\"u}li, and {\relax Th}.~P. Wihler.
\newblock A posteriori error analysis of {$hp$}-version discontinuous
  {Galerkin} finite-element methods for second-order quasi-linear elliptic
  {PDE}s.
\newblock {\em {IMA J. Numer. Anal.}}, 28(2):245--273, 2008.

\bibitem{Kloefkorn2009}
R.~Kl{\"o}fkorn.
\newblock {\em Numerics for evolution equations: {A} general interface based
  design concept}.
\newblock Doctoral dissertation, Albert-Ludwigs-Universit{\"a}t Freiburg, 2009.

\bibitem{Lage1998}
C.~Lage.
\newblock Concept oriented design of numerical software.
\newblock Research Report 98-07, Seminar for Applied Mathematics, ETH
  Z{\"u}rich, 1998.

\bibitem{Mitchell2006}
W.~F. Mitchell.
\newblock {PHAML} user's guide.
\newblock Technical Report NISTIR 7374, National Institute of Standards and
  Technology, 2006.

\bibitem{Mitchell2011}
W.~F. Mitchell and M.~A. McClain.
\newblock A comparison of {$hp$}-adaptive strategies for elliptic partial
  differential equations (long version).
\newblock Technical Report 7824, National Institute of Standards and
  Technology, 2011.

\bibitem{Mitchell2014}
W.~F. Mitchell and M.~A. McClain.
\newblock A comparison of {$hp$}-adaptive strategies for elliptic partial
  differential equations.
\newblock {\em {ACM Trans. Math. Softw.}}, 41(1):2:1--2:39, 2014.

\bibitem{Schupp1999}
B.~Schupp.
\newblock {\em {Entwicklung} eines effizienten {Verfahrens} zur {Simulation}
  kompressibler {Str{\"o}mungen} in {3D} auf {Parallelrechnern}}.
\newblock Doctoral dissertation, Albert-Ludwigs-Universit{\"a}t Freiburg, 1999.

\bibitem{Solin2004}
P.~{\v{S}}ol{\'i}n and L.~Demkowicz.
\newblock Goal-oriented {$hp$}-adaptivity for elliptic problems.
\newblock {\em {Comput. Method. Appl. M.}}, 193(6--8):449--468, 2004.

\bibitem{Zhu2011}
L.~Zhu, S.~Giani, P.~Houston, and D.~Sch{\"o}tzau.
\newblock Energy norm a posteriori error estimation for {$hp$}-adaptive
  discontinuous {Galerkin} methods for elliptic problems in three dimensions.
\newblock {\em {Math. Models Methods Appl. Sci.}}, 21(02):267--306, 2011.

\end{thebibliography}
\end{document}